\documentclass[acmsmall]{acmart}

\usepackage{multirow}
\usepackage{caption}
\usepackage{subcaption}
\usepackage{enumitem}
\usepackage[normalem]{ulem}
\usepackage{url}

\AtBeginDocument{%
  }

\setcopyright{cc}
\setcctype{by-nc}
\acmJournal{PACMHCI}
\acmYear{2025} \acmVolume{9} \acmNumber{7} \acmArticle{410} \acmMonth{11} \acmPrice{}\acmDOI{10.1145/3757591}

\received{October 2024}
\received[revised]{April 2025}
\received[accepted]{August 2025}

\newcommand{\parabold}[1]{\noindent\textbf{#1.}}

\definecolor{change}{RGB}{0,0,0}
\definecolor{changes}{RGB}{0,0,0}
\begin{document}

\title{Examining Human-AI Collaboration for Co-Writing Constructive Comments Online}
\author{Farhana Shahid}
 \affiliation{
   \institution{Cornell University}
   \city{Ithaca}
   \country{United States}
   }
\email{fs468@cornell.edu}

\author{Maximilian Dittgen}
 \affiliation{
   \institution{Cornell University}
   \city{Ithaca}
   \country{United States}
   }
\email{myd4@cornell.edu}

\author{Mor Naaman}
 \affiliation{
   \institution{Cornell Tech}
   \city{NYC}
   \country{United States}
   }
\email{mor.naaman@cornell.edu}

\author{Aditya Vashistha}
 \affiliation{
   \institution{Cornell University}
   \city{Ithaca}
   \country{United States}
   }
\email{adityav@cornell.edu}


\begin{abstract}

This paper examines if large language models (LLMs) can help people write constructive comments on divisive social issues due to the difficulty of expressing constructive disagreement online. Through controlled experiments with 600 participants from India and the US, who reviewed and wrote constructive comments on threads related to Islamophobia and homophobia, we observed potential misalignment between how LLMs and humans perceive constructiveness in online comments. While the LLM was more likely to prioritize politeness and balance among contrasting viewpoints when evaluating constructiveness, participants emphasized logic and facts more than the LLM did. Despite these differences, participants rated both LLM-generated and human-AI co-written comments as significantly more constructive than those written independently by humans. Our analysis also revealed that LLM-generated comments integrated significantly more linguistic features of constructiveness compared to human-written comments. When participants used LLMs to refine their comments, the resulting comments were more constructive, more positive, less toxic, and retained the original intent. However, LLMs often distorted people's original views---especially when their stances were on a spectrum instead of being outright polarizing. Based on these findings, we discuss ethical and design considerations in using LLMs to facilitate constructive discourse online.


\end{abstract}



\maketitle

\section{Introduction}


Most people use social media in good faith~\cite{rajadesingan2021} but struggle to reach common ground during online conflicts, which often lead to toxicity and personal attacks~\cite{Baughan2021, gurgun2023}. The lack of support for constructive disagreement online discourages people from challenging problematic content~\cite{gurgun2023, shahid2024one}. This results in downstream harms, such as the disappearance of minority viewpoints~\cite{Collier2012, Grevet2014} and increased propagation of harmful content~\cite{shahid2024one}.

Researchers define \textbf{constructive} comments as those that balance argumentation with politeness~\cite{Kolhatkar2017a, kolhatkar2017b, kolhatkar2020}. 
HCI researchers have explored several 
strategies to promote healthy discourse, such as using social cues to highlight positive behavior~\cite{jhaver2017, rajadesingan2021}, and introducing frictions~\cite{Seering2019, katsaros2022, Masrani2023} and nudges~\cite{esau2017, Taylor2019, bossens2021, park2023} to discourage people from using offensive language. However, these interventions put the onus on users to write their opinions constructively, which is already difficult in non-conflicting situations~\cite{Cutler2022}. 
Since crafting constructive responses can be time-consuming and cognitively demanding, people are often discouraged from expressing disagreement online~\cite{Mun2024, gurgun2023}. 

To address this,
researchers have begun training large language models (LLMs) to help people in argumentative writing~\cite{Lee2022, Zhang2023, Dang2023}. 
While early evaluations show that LLMs can improve people's argument~\cite{Zhang2023} and help them find common ground on divisive issues~\cite{argyle2023}, much of this work focuses on people's experience of writing argumentative essays or messages in private debates. Additionally, these studies overlook cultural differences in argumentation, which is important because while expressing disagreement online, people expect writing support that would reflect their socio-cultural norms~\cite{Mun2024}.

Prior research shows that people from individualistic cultures prefer \textit{logical} arguments that follow formal rules to substantiate one's claims~\cite{nisbett2001culture, Peng1999, choi1988}. In contrast, people from collectivist cultures favor \textit{dialectical} arguments during social conflicts, emphasizing a holistic consideration of different viewpoints and finding middle ground~\cite{ting1991, ting1998, Peng1999}. Since argumentation is a key element of constructive comments~\cite{Kolhatkar2017a, kolhatkar2017b}, it is crucial to consider if cultural differences in argumentation affect people's perceptions of constructiveness---especially when when designing interventions to support constructive discourse on divisive social issues. While LLMs show promise in supporting argumentative writing~\cite{Zhang2023}, these models often homogenize writing to reflect Western norms, erasing important cultural nuances~\cite{agarwal2024}. This raises the need to assess whether perceptions of constructiveness align between humans and LLMs, particularly across culturally varied argumentation styles. Therefore, our study explores two key research questions:

\begin{itemize}
    \item[\textbf{RQ1:}] Do perceptions of constructiveness vary among 
    humans and LLMs based on the argumentation style?
    \item[\textbf{RQ2:}] Can LLMs help people from different cultures write constructive comments on divisive issues?
\end{itemize}

To answer these questions, we conducted a two phase study with participants from India and the US, who reviewed homophobic and Islamophobic threads relevant to their cultural contexts.

\textbf{In Phase 1}, we used GPT-4 to generate constructive comments in response to these threads, following either a \textit{logical} or \textit{dialectical} argumentation style. 
We used a combination of qualitative crowd evaluation and quantitative analyses of linguistic features to verify that the LLM-generated comments varied in argumentation style while maintaining similar levels of constructiveness across different linguistic features. Next, we conducted a forced-choice experiment with 103 Indian and American crowd workers on Prolific. Each participant reviewed either a homophobic or an Islamophobic thread, along with randomly selected pairs of logical vs. dialectical comments, written for the same thread from the same stance. 
For each pair, participants indicated which comment they perceived as more constructive and why. To compare perceptions of constructiveness between humans and LLMs, we also assigned the same task to GPT-4.

In response to \textbf{RQ1}, we found that both Indian and American participants and GPT-4 perceived dialectical comments as more constructive than the logical ones. 
However, GPT-4 was 2.46 times more likely than humans to select dialectical comments as more constructive than the logical ones. We found that GPT-4 highlighted politeness and the inclusion of contrasting viewpoints---hallmarks of dialectical argument---when assessing constructive comments. In contrast, participants placed greater emphasis on logical structure and facts, features of logical argumentation, than GPT-4 did. 
These findings point to a potential misalignment between how humans and LLMs assess what makes a comment constructive. 

\textbf{In Phase 2}, to examine how LLMs facilitate constructive writing, we conducted a between-subject experiment with a different set of 103 Indian and American crowd workers on Prolific. Participants reviewed homophobic and Islamophobic threads and wrote constructive comments on these threads. Participants were randomly assigned to either a control group where they wrote comments independently, or a test group where they could select prompts to request an LLM to rewrite their comments constructively. In the test group, participants had the flexibility to accept, edit, reject, or regenerate the suggestions from LLM.

In response to \textbf{RQ2}, we compared the comments participants wrote independently in control group (human-written),  or with LLM's assistance in test group (human-AI written)---alongside the comments solely generated by LLM (AI) in Phase  1. 
Crowd evaluation showed that, when presented with (Human vs. AI) comment pairs, participants were 8.51 times more likely to select LLM-generated comments as more constructive than the human-written comments. Similarly, participants were 3.19 times more likely to choose HAI-written comments as more constructive over human-written comments. Quantitative analyses revealed that LLM-generated comments contained significantly more linguistic features of constructiveness, such as greater length, better readability, and more argumentative features than human-written comments. %

We also found that when participants used LLM to rewrite their comments, it made their comments significantly more positive, more constructive, and less toxic. In most cases, participants accepted the LLM's suggestions because it conveyed their points better without homogenizing their writing. However, some participants felt that the LLM misrepresented their views and they edited LLM's suggestions in ways that made their comments more negative and toxic. Overall, most participants were satisfied with the comments they wrote using LLMs and found the process was easier than writing constructive comments independently. 

With respect to cultural differences, although prior work highlights cross-cultural variation in argumentative writing~\cite{liu2005rhetorical, liu2009}, we found no significant differences in the linguistic markers of constructiveness in comments written by Indian and American participants. This suggests a shared understanding of constructiveness across different cultures, particularly in the case of expressing disagreements online on divisive social issues. Taken together, our work makes the following contributions:

\begin{itemize}[leftmargin=1.75\parindent]
    \item We provide both quantitative and qualitative evidence that LLMs can help people from different cultures write constructive comments on divisive issues.
    \item We uncover potential misalignment between how humans and LLMs characterize constructiveness prioritizing different argumentation styles. 
    \item We reveal the potential risk of LLMs misrepresenting people's views on divisive issues by applying positive sentiment to their comments.
    \item We discuss both ethical and design consideration for developing socio-technical systems that promote constructive discourse on divisive issues across different cultures.
\end{itemize}
\section{Related Work}

Researchers have designed various interventions to promote civil dialogue online. Research shows that user-level strategies, such as highlighting top community members~\cite{jhaver2017} or shared interests~\cite{rajadesingan2021}, encourage politeness. Platform-level measures, such as listing group rules~\cite{kraut2012building} and allowing private discussions~\cite{Baughan2021, kim2022}, help reduce conflicts. Comment-level approaches, including highlighting toxicity~\cite{warner2024} and prompting reflection before posting~\cite{Taylor2019, Masrani2023, park2023, katsaros2022, Seering2019} lead to empathetic comments. Structured support, such as offering discussion points~\cite{bossens2021} and specific questions~\cite{esau2017}, promotes civil exchange. Thread-level interventions, like surfacing high-quality comments~\cite{Berry2017, Wang2022}, signaling tension~\cite{Chang2022}, and summarizing key points~\cite{Kriplean2012, Kriplean2012a}, improve discussion quality.

While such interventions aim to promote prosocial behavior online, research shows that users often find it emotionally and cognitively taxing to respond constructively~\cite{Mun2024, gurgun2023}. They often prefer pre-written questions~\cite{gurgun2023} and sentence openers~\cite{Mcalister-2004} that help them challenge others constructively instead of having to write from scratch. In this context, recent advances in large language models (LLMs) have opened up new avenues to support argumentative writing. We situate our work first by discussing the role of LLMs in supporting users during online disagreements. We then discuss prior research on constructive disagreement, along with cross-cultural variations in argumentation.

\subsection{LLMs for Facilitating Online Argumentation}

Several researchers have used LLMs as mediators to improve online argumentation. For example, \citet{Govers2024} conducted an experiment with American participants who reviewed polarizing online threads containing comments from both human and LLM-based mediators. They found that highly cooperative and persuasive strategies deployed by mediator-bots were effective in changing reader's opinions. 
Similarly, \citet{Tessler-2024} fine-tuned an LLM 
to craft opinion statements on divisive political issues in the UK
and found that it helped small groups find common ground. 

Beyond mediating divisive issues, LLMs have been used to provide users with actionable feedback during writing. For instance, \citet{Zhang2023} developed a tool that helped writers visualize and integrate elements of logical argumentation--such as claim, data, warrant, backing, qualifier, and rebuttal---into their writing. They evaluated the tool among US college students who found the tool helpful for writing argumentative essays. Similarly, \citet{Xia2022} designed an interactive visual system that highlighted which persuasive strategies (i.e., logos, pathos, ethos, and evidence) were present or missing in users' response, helping non-English speaking users write persuasive arguments in English.~\citet{ding2024} developed a learning tool for the US-based native English speakers, which helped people brainstorm counter-speech strategies and guided them to use empathetic tone while challenging hate speech. 
In another study, \citet{argyle2023} conducted an experiment with American participants in which an LLM prompted users to rephrase their message either by making them more polite, restating opposition’s arguments, or validating opposition’s sentiment. Their findings show that rephrasing messages to demonstrate respectful listening in one-on-one debates improved conversation quality, openness to opposing views, and participants' sense of being understood. Finally, \citet{kambhatla2024} curated human-written comments on controversial topics and reframed the comments using LLM to incorporate receptive elements, such as hedging, acknowledgment, 
or agreement. They found that US participants perceived the LLM-rephrased versions as more receptive to opposing views than the original comments.

While these studies employ diverse approaches to improve discussion on divisive issues, very few actually examine users' experiences of engaging in argumentative writing. The handful of studies using LLMs to facilitate argumentative writing either focus on argumentative essays~\cite{Lee2022, Zhang2023, Dang2023}, private one-on-one debates~\cite{argyle2023}, or static learning environments~\cite{ding2024}. These contexts are qualitatively different from responding to an online thread on divisive issues, which often draws hateful interactions from users~\cite{napoles-2017}. Moreover, existing research on LLM-based interventions primarily focuses on Western populations, leaving a significant gap in understanding how to foster constructive discourse across different cultures~\cite{shortall2021reason}. This is particularly important because most online users come from non-Western regions~\cite{Knowledge-2022}. Yet, LLMs tend to prioritize Western norms and values~\cite{cao2023, johnson2022}, often homogenizing writings to align with Western styles~\cite{agarwal2024}. 

To address these gaps, we examine if LLMs can facilitate constructive discourse on divisive issues in different cultural settings. In doing so, we specifically focus on \textit{``constructiveness''}, as online platforms routinely moderate and rank users' comments based on constructiveness~\cite{Diakopoulos2011, diakopoulos2015} and often explicitly guide users to be constructive~\cite{ludwig2014}. To this end, we build upon prior work on constructive discourse and cross-cultural differences in argumentation, which we discuss next.

\subsection{Constructiveness in Online Discussion}
\label{sec:constructiveNLP}

\parabold{Subjective Interpretation} The concept of constructiveness varies depending on the context. In student evaluations, peer reviews, and product reviews, constructive criticism refers to respectful feedback that offers actionable suggestions for improvement~\cite{fong2016, weaver2022art, yan2018posters}. In cases of disagreement, constructive conflict resolution involves demonstrating cooperation and building trust with opposing parties~\cite{deutsch1994constructive, bachtiger2019mapping}. 

In online discussions, constructiveness takes on additional nuances. \citet{friess2015} define constructive deliberation as finding common ground and proposing new solutions. \citet{Kolhatkar2017a} surveyed people who described constructive comments as civil dialogues that are relevant, address specific points, and provide appropriate evidence. Other studies define constructive disagreements based on their outcomes, such as whether the dispute is resolved~\cite{de-kock-vlachos-2021} or leads to improved team performance~\cite{mizil-2016-conversational}. While these definitions emphasize the subjective nature of what is considered constructive, researchers in Natural Language Processing (NLP) have tried to capture the linguistic features of constructiveness, as outlined below.

\parabold{Linguistic Features} To detect constructive features, NLP researchers have analyzed online conversations from different sources---such as CNN~\cite{Sukumaran2011}, New York Times~\cite{Sukumaran2011, kolhatkar2017b}, Yahoo News~\cite{napoles-2017, Kolhatkar2017a}, online games~\cite{mizil-2016-conversational}, and Wikipedia~\cite{zhang-etal-2018, de-kock-vlachos-2021}---and relied on human evaluations to annotate these conversations along different dimensions, such as tone, level of agreement, and constructiveness. They also 
extracted different linguistic features (e.g., toxicity, politeness) from these conversations, 
and trained classifiers to identify significant predictors of constructiveness.

Findings from NLP research show that constructive comments tend to be more issue-relevant \cite{Sukumaran2011} and contribute to the main points in the conversation~\cite{kolhatkar2017b}. They are usually longer in length and take longer to write~\cite{Sukumaran2011, kolhatkar2017b}. Constructive comments are more likely to show disagreement and thus, contain less hedging (less hesitation) and matched language (less subordinate)~\cite{mizil-2016-conversational, napoles-2017, de-kock-vlachos-2022-disagree}. They are more likely to contain argumentative features, such as discourse connectives, stance adverbials, reasoning verbs and modals, and root clauses~\cite{Kolhatkar2017a, napoles-2017}. Additionally, respectful attitude and different politeness strategies, such as gratitude, greetings, requests, and deference are often observed in constructive discourse~\cite{napoles-2017, zhang-etal-2018, de-kock-vlachos-2021}. Readability scores, presence of solutions, evidence, personal stories~\cite{kolhatkar2017b}, and named entities~\cite{kolhatkar2020, mizil-2016-conversational} are also indicative of constructiveness.

These findings suggest that constructive discourse requires balancing argumentation with politeness to express disagreements assertively, without appearing hesitant or subordinate. However, these features are mostly derived from discussions in Western media outlets and the quality of comments (e.g., respectfulness, reasoning) on online news websites vary a lot depending on the country's attitude towards deliberation~\cite{ruiz2011public}. Therefore, we look into the cross-cultural differences in how people approach argumentation during social conflicts, which we discuss next.

\subsection{Cross-Cultural Differences In Argumentation}
\label{sec:culturalDiff}
Research shows that argumentation styles vary across individualistic (e.g., Western countries) and collectivist cultures (e.g., Asian countries). People from collectivist cultures like China~\cite{xie2015cross} and India~\cite{hample2015understandings} often view face-to-face arguments as less civil and more personal, and report feeling persecuted during interpersonal disagreements. Due to higher power distance within such cultures, people are more likely to avoid dissent or engage in it strategically, in contrast to their American counterparts~\cite{kapoor2003, croucher2009}. They tend to be more polite, indirect, and moderate in expressing disagreement~\cite{hill1986, holtgraves1990, ting1991, Peng1999, norenzayan2002cultural, suzuki2010}.

These cultural differences also influence reasonoing styles. People from collectivist cultures often engage in \textit{dialectical} argumentation---an approach that embraces contradiction and seeks balance---while those from individualistic cultures typically prefer \textit{logical}  reasoning, which emphasizes formal rules and binary outcomes~\cite{norenzayan2000rules, norenzayan2002cultural}. When faced with conflicting viewpoints, Americans and Europeans tend to take polarized positions, aiming to determine which fact and position are right or wrong~\cite{Peng1999, nisbett2001culture}. This results in argumentation that is more direct, assertive, solution-oriented, and dominant~\cite{ting1991, ting1998}. Therefore, argumentative essays written by American students are more direct, assertive, and apply formal logic, where each claim is followed by supporting evidence~\cite{choi1988, liu2005rhetorical}. In contrast, dialectical argumentation assumes no claim is absolute; rather every viewpoint is contextually bound. 
As a result, people from collectivist cultures tend to holistically consider the relations among different perspectives and are more willing to take a ``middle ground'' between conflicting viewpoints~\cite{peng1997naive, Peng1999, nisbett2001culture, norenzayan2002cultural, choi2007individual, hample2015understandings}. 
In writing, this manifests as developing arguments by incorporating contrasting viewpoints~\cite{wu2000evaluating, liu2005rhetorical}. See Table~\ref{tab:con_comments} for illustrative examples.

While much of this research focuses on offline conflict and formal argumentative writing, online interactions differ in key ways. 
For example, online disinhibition effect allows people to use harsher language, express stronger criticism, and behave in less socially acceptable ways online than they do offline~\cite{suler2004online}. 
Although cross-cultural research on online conflict is limited, the findings are mixed: 
\citet{fichman2022impact} found that Americans tend to write more trolling comments than Indians, whereas \citet{cook2021trolls} found that Taiwanese tend to react more aggressively than the Dutch to save their face when instigated online. These results suggest that cultural norms continue to shape argumentative behavior online, though not always predictably.

Despite the importance of argumentation in shaping constructive discourse, so far no systematic study has examined how constructive disagreement varies across cultures. To address this gap—and in light of the support people need to express disagreement online—we conducted a cross-cultural study with participants from India and the US. We first investigate if perceptions of constructive comments differ based on different argumentation styles (RQ1). We then examine if LLMs could help people from different cultures write constructive comments on divisive issues (RQ2). 

\section{Methods}

\begin{figure}[ht]
    \centering
    \includegraphics[width=0.9\linewidth, trim={.7cm 1cm .8cm 1cm}, clip]{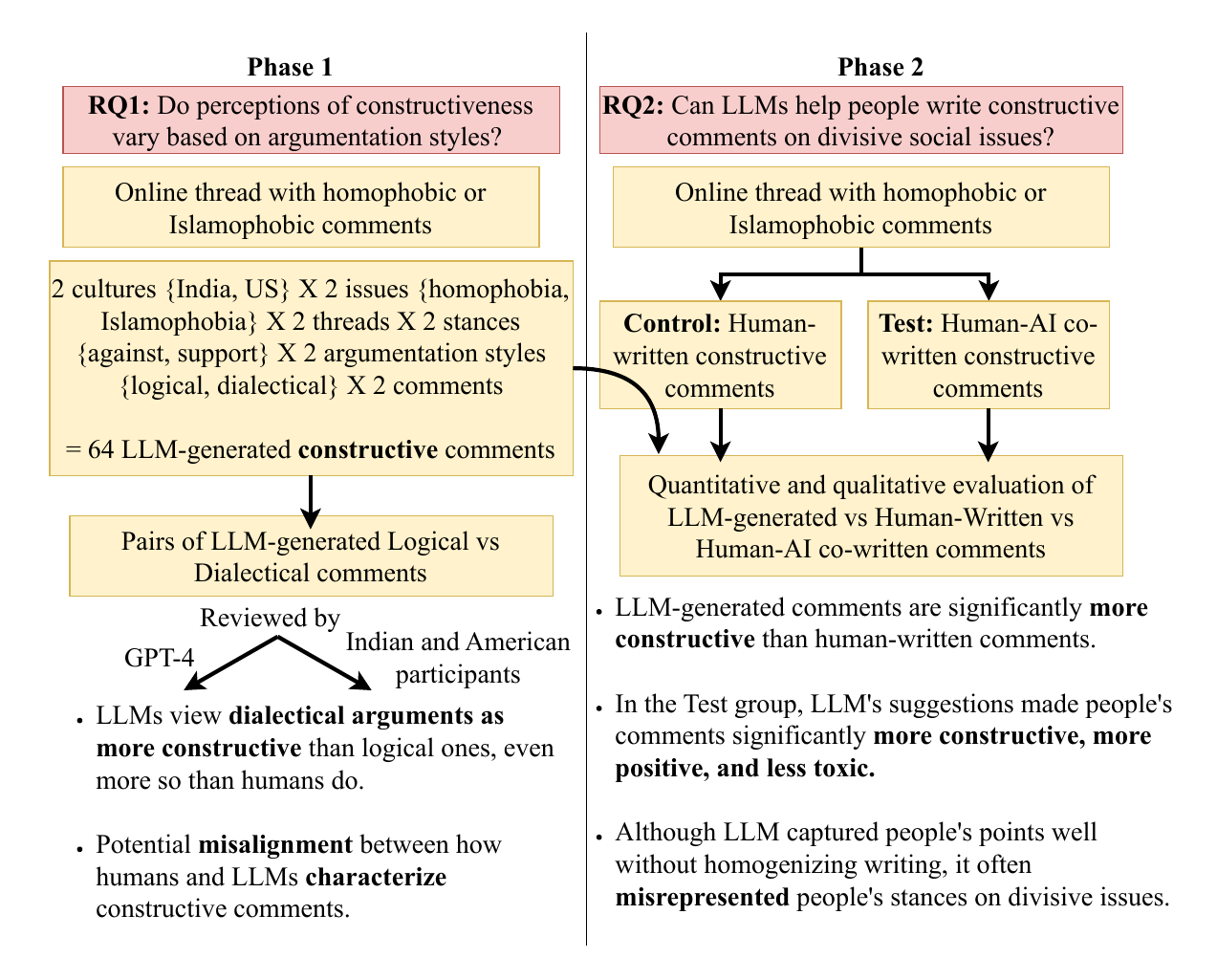}
    \caption{Flow chart of our two-phase study investigating perceptions and writing of constructive comments.}
    \label{fig:flowchart}
    \Description{Flow chart of our two-phase study investigating perceptions and writing of constructive comments.}
\end{figure}

To address our research questions, we conducted a two-phase study (see Figure~\ref{fig:flowchart}) with participants from both individualistic (the US) and collectivist cultures (India). In the first phase, we examined if perceptions of constructive comments on socially divisive issues vary depending on different argumentation styles present in writing. In the second phase, we asked people to write constructive comments either with or without the help of an LLM. The study received exemption from institutional review board (IRB) at our institution.

\subsection{Phase 1: Perceptions of Constructive Comments}
Since prior research indicates that people from individualistic cultures prefer logical arguments, while those from collectivist cultures favor dialectical arguments, we examined in Phase 1 if the perception of constructiveness varies depending on the argumentation style. 

\parabold{Selection of Topics and Online Threads} For our study, we selected Reddit threads related to homophobia and Islamophobia because these issues are considered socially divisive in both the US and India. For each country, we curated two homophobic and two Islamophobic threads (see Table \ref{tab:thread_titles} in Appendix), each consisting of the original post and four user comments from relevant subreddits. 
The threads for American participants were drawn from r/conservative, r/politics, r/atheism, and r/changemyview, and the threads for Indian participants were sourced from r/IndiaSpeaks. All threads were in English and 78\% of comments on these threads were negative in sentiment (average: -0.63) and included several highly negative and toxic comments.  

\parabold{Generating Constructive Comments with Different Argumentation Styles} Next, we used GPT-4 to generate constructive comments on the selected threads for each country. Since homophobia and Islamophobia are divisive issues, we wanted to take into account participants' personal stance on these topics when they would evaluate constructiveness. We thus generated comments for two different stances---$\{\mathrm{against, supportive}\}$---because these capture the two definitive and polarizing stances of people when they engage with such divisive topics. Considering the cultural orientation, we generated constructive comments following two different argumentation styles: $\{\mathrm{logical, dialectical}\}$. In total, we generated comments for 2 cultural contexts $\mathrm{\{India, US\}}$ $\times$ 2 issues $\mathrm{\{homophobia, Islamophobia\}}$ $\times$ 2 threads $\times$ 2 stances $\mathrm{\{against, support\}}$ $\times$ 2 argumentation styles $\mathrm{\{logical, dialectical\}}$ = 32 different cases. 

To ensure that participants' responses are not sensitive to particular wordings, we first generated three different comments for each case, resulting in $32 \times 3=96$ comments in total (48 for the US, 48 for India). A subset of these comments were used in the study, after validation by human annotators. To generate these comments, we used zero-shot, cultural prompting (i.e., specified the country) to increase cultural alignment in LLM-generated comments for both countries~\cite{tao2024}. See Appendix for more details on model hyper-parameters. We instructed GPT-4 to keep the comments within 100 words, which is the average length of constructive comments as observed in prior study~\cite{kolhatkar2017b}. Table~\ref{tab:con_comments} shows examples of LLM-generated constructive comments with different argumentation styles, and Table \ref{tab:comment_gen} in the Appendix shows the prompts used to generate constructive comments. 

\begin{table}
\caption{Examples of LLM-generated constructive comments with different argumentation styles. \colorbox{Salmon}{Pink colored segments} demonstrate how logical arguments take direct, polarizing stance followed by supporting examples and claims. \colorbox{SpringGreen}{Green colored segments} show how dialectical arguments acknowledge contrasting viewpoints and take a middle-ground when developing arguments.}
\label{tab:con_comments}
\resizebox{\columnwidth}{!}{

\begin{tabular}{l|l|l|}
\cline{2-3}
\multicolumn{1}{c|}{}                                                                                                                      & \multicolumn{1}{c|}{\textbf{Logical argumentation}}                                                                                                                                                                                                                                                                                                                                                                                                                                                                                                                                                                                              & \multicolumn{1}{c|}{\textbf{Dialectical argumentation}}                                                                                                                                                                                                                                                                                                                                                                                                                                                                                                                                                                                                          \\ \hline
\multicolumn{1}{|l|}{\begin{tabular}[c]{@{}l@{}}Demographic: USA\\ Issue: Homophobia\\ Stance: Supports \\ same-sex marriage\end{tabular}} & \begin{tabular}[c]{@{}l@{}}The \colorbox{Salmon}{assumption that LGBTQ+} \\ \colorbox{Salmon}{individuals are merely seeking }\\ \colorbox{Salmon}{victimhood ignores systemic} \\ \colorbox{Salmon}{legal and social challenges} they \\ face. Laws restricting their rights, \\ such as \colorbox{Salmon}{bans on gender-affirming} \\ \colorbox{Salmon}{care or anti-LGBTQ+ rhetoric} in \\ politics, \colorbox{Salmon}{have real consequences,} \\ not imagined grievances. \colorbox{Salmon}{Dismissing}\\ \colorbox{Salmon}{their concerns as "attention-seeking"} \\ \colorbox{Salmon}{is an oversimplification} that ignores \\ historical patterns of discrimination \\ and marginalization. \colorbox{Salmon}{Protecting equal} \\ \colorbox{Salmon}{rights is} not about forcing beliefs but \\ \colorbox{Salmon}{ensuring legal fairness for all.}\end{tabular}                            & \begin{tabular}[c]{@{}l@{}}Discussions about LGBTQ+ rights \\ often reflect \colorbox{SpringGreen}{deeper societal tensions} \\ \colorbox{SpringGreen}{between tradition and modern values.}\\ While \colorbox{SpringGreen}{some feel that advocacy has} \\ \colorbox{SpringGreen}{become overly forceful, others see it} \\ \colorbox{SpringGreen}{as a necessary response to real} \\ \colorbox{SpringGreen}{struggles.} Perhaps the key lies in \\ \colorbox{SpringGreen}{balancing personal freedoms with} \\ \colorbox{SpringGreen}{respectful dialogue,} ensuring that \\ \colorbox{SpringGreen}{neither side feels silenced} nor \\ threatened. In any case, \colorbox{SpringGreen}{mutual} \\ \colorbox{SpringGreen}{understanding} may be more \\ productive than\colorbox{SpringGreen}{ viewing each} \\ \colorbox{SpringGreen}{other as opponents.}\end{tabular}                                                                                                               \\ \hline
\multicolumn{1}{|l|}{\begin{tabular}[c]{@{}l@{}}Demographic: India\\ Issue: Islamophobia\\ Stance: Critical of\\ Islam\end{tabular}}       & \begin{tabular}[c]{@{}l@{}}Honestly, it baffles me how some \\ folks dismiss historical facts. Frankly, \\ \colorbox{Salmon}{Islamic invaders did destroy countless} \\ \colorbox{Salmon}{Hindu temples} during their reign, it's \\ irrefutable. Imagining the brutal \\ persecution Hindus must have \\ endured is chilling. Manifestly, \\ \colorbox{Salmon}{the issue isn't about coexistence,} \\ \colorbox{Salmon}{because other religions in India} \\ \colorbox{Salmon}{have managed to do so.} It's the \\ \colorbox{Salmon}{extremist interpretation of Islam} \\ \colorbox{Salmon}{that's problematic.} Evidently, the \\ \colorbox{Salmon}{onus lies on Islamic society to} \\ \colorbox{Salmon}{reform and eradicate such bigotry,} \\ lest we remain trapped in this \\ vicious cycle of hatred.\end{tabular} & \begin{tabular}[c]{@{}l@{}}\colorbox{SpringGreen}{While I understand the historical} \\ \colorbox{SpringGreen}{grievances,} it's important to \\ \colorbox{SpringGreen}{distinguish between Islamist} \\ \colorbox{SpringGreen}{extremists and everyday Muslims.} \\ Extremism exists in all religions \\ and is always harmful. Remember, \\ \colorbox{SpringGreen}{many Muslims in India also have a} \\ \colorbox{SpringGreen}{long ancestry here, so they're as} \\ \colorbox{SpringGreen}{native as us.} Let's aim for \\ \colorbox{SpringGreen}{coexistence, acknowledging} \\ \colorbox{SpringGreen}{past atrocities} but focusing on \\ preventing future ones. \colorbox{SpringGreen}{Rather} \\ \colorbox{SpringGreen}{than being anti-Muslim, we should} \\ \colorbox{SpringGreen}{be anti-extremism} and strive for a \\ harmonious society. We all have \\ \colorbox{SpringGreen}{a shared responsibility} to ensure \\ peace and unity in our diverse nation.\end{tabular} \\ \hline
\end{tabular}
}
\end{table}

\parabold{Validating Argumentation Style} 
To ensure that LLM-generated comments indeed followed logical or dialectical argumentation styles, we conducted a validation study with 230 crowd workers from MTurk in India and the US. We first asked annotators about their views on same-sex marriage and Islam, using adapted instruments from Pew research survey~\cite{Borelli-2024, pew-2024}. Then, we showed them either one randomly selected homophobic or Islamophobic thread, along with four randomly selected LLM-generated comments written for that thread. We presented annotators with comments that matched their stance on homophobia or Islamophobia, to account for biases from annotators' personal opinions on these issues. For annotators, who reported feeling neutral about the issues, we randomly showed them comments that either supported or opposed the issue.

We asked participants to annotate if a given comment either followed logical step-by-step arguments (logical) or holistically considered different viewpoints and took a middle-ground (dialectical). Each annotator reviewed four comments and received $\$1.00$ as compensation. 

Once we gathered at least five annotations for each comment, we calculated the agreement in annotations. For comments that were labeled incorrectly or where the annotators did not reach agreement, we generated new comments and repeated the process. We continued this process until we had at least two comments for each of the 32 cases, where the majority annotation (60\% or above) matched the actual labels (See Table \ref{tab:qual_eval_arg}). In the end, we finalized 64 comments (US: 32, India: 32), where the argumentation styles in the LLM-generated comments were validated by human annotators.   
\begin{table}
\caption{Validation of the argumentation styles in LLM-generated comments.}
\label{tab:qual_eval_arg}
\resizebox{\columnwidth}{!}{
\begin{tabular}{|c|c|c|c|c|}
\hline
Country      & \begin{tabular}[c]{@{}c@{}}Number of\\ annotators\end{tabular} & \begin{tabular}[c]{@{}c@{}}Number of annotated\\ comments\end{tabular} & \begin{tabular}[c]{@{}c@{}}Number of final comments where \\ majority annotation matched true label\end{tabular} & \begin{tabular}[c]{@{}c@{}}Annotations per\\ final comment\end{tabular} \\ \hline
US   & 142                                                            & 66                                                                     & 39                                                                                                               & Avg: 7.5 (SD: 3.6)                                                      \\ \hline
India & 88                                                             & 57                                                                     & 32                                                                                                               & Avg: 5.9 (SD: 2)                                                        \\ \hline
\end{tabular}}
\end{table}

\parabold{Assessing Level of Constructiveness} 
To verify if LLM-generated comments differed beyond argumentation styles (logical vs. dialectical), we analyzed the key linguistic features of constructiveness as reported in prior work~\cite{kolhatkar2017b, napoles-2017, zhang-etal-2018}. These are: comment length, readability score, presence of politeness strategies, named entities, and argumentative features, such as discourse connectives, stance adverbials, reasoning verbs and modals, and root clauses (see Table~\ref{tab:constructive-comment-feature}).

\begin{table}[t]
\caption{Linguistic features of constructiveness as reported in prior work~\cite{kolhatkar2017b, napoles-2017, zhang-etal-2018}}
\label{tab:constructive-comment-feature}
\resizebox{\columnwidth}{!}{
\begin{tabular}{|l|l|l|}
\hline
\textbf{Features}                                                                 & \textbf{Association with constructiveness}                                                                                                                                                                                                        & \textbf{Measure}                                                                \\ \hline
Length                                                                            & Longer in length                                                                                                                                                                                                                                  & Word count                                                                      \\ \hline
Readability score                                                                 & High levels of readability                                                                                                                                                                                                                        & \begin{tabular}[c]{@{}l@{}}SMOG index~\cite{park2023}\\ \textit{textstat} library in Python\end{tabular} \\ \hline
Politeness                                                                        & Respectful and uses different politeness strategies                                                                                                                                                                                               & \textit{politeness} package in R~\cite{yeomans2018}                                                         \\ \hline
Named entities                                                                    & Contains named entities                                                                                                                                                                                                                           & SpaCy~\cite{explosion2017spacy}                                                                           \\ \hline
\multirow{2}{*}{\begin{tabular}[c]{@{}l@{}}Argumentative\\ features\end{tabular}} & \begin{tabular}[c]{@{}l@{}}Greater presence of\\ 1. discourse connectives (e.g., therefore, due to)\\ 2. stance adverbials (e.g., of course, undoubtedly)\\ 3. reasoning verbs (e.g., cause, lead), and\\ modals (e.g., may, should)\end{tabular} & \begin{tabular}[c]{@{}l@{}}\textcolor{changes}{Adapted open-source} \\ \textcolor{changes}{code from \citet{kolhatkar2020}}\end{tabular}        \\ \cline{2-3} 
                                                                                  & 4. root clauses (e.g., I think that)                                                                                                                                                                                                              & \begin{tabular}[c]{@{}l@{}}\textcolor{changes}{dependency parsing} \\ \textcolor{changes}{from SpaCy}~\cite{explosion2017spacy}\end{tabular}        \\ \hline
\end{tabular}}
\end{table}


We found that LLM-generated logical and dialectical comments were similar across several linguistic markers of constructiveness, such as length, politeness, presence of named entities, discourse connectives, and the use of reasoning verbs and modals (see Table~\ref{tab:constructive-feature-ai-gen} in the Appendix). However, dialectical comments were significantly more readable, while logical comments featured significantly more stance adverbials and root clauses.

\parabold{Experiment 1: Capturing Perceptions of Constructiveness} 
To examine if the notions of constructiveness differ depending on argumentation styles (RQ1), we conducted a forced-choice experiment with 103 crowd workers from Prolific (US: 51, India: 52). In the experiment, participants were presented with a randomly selected homophobic or Islamophobic thread. The comments in the thread appeared in random order with redacted user names and profile photos (see Figure~\ref{fig:us-islamophobic-post} in the Appendix). Given the toxicity in these threads, we warned the participants that they might get exposed to negative comments before they accepted the task and gave them the option to quit the study at any point if they felt uncomfortable engaging with such content. 

The participants, who accepted the task, reviewed four pairs of comments--where each pair contained one randomly selected logical and one dialectical comment generated by LLM for the same thread. Participants saw each comment only in one pair, and the order of comments within each pair was randomized. Among the four pairs, two pairs consisted of comments opposing the issue, while the other two pairs consisted of comments supporting the issue. This way, participants were exposed to comments that did not align with their personal stances, as is common in online platforms. Since both comments within a pair shared the same stance, participants were not biased toward selecting the comment that matched their views.

We then asked participants to indicate which comment they perceived as more constructive in each pair and why. We gave them several options adapted from the characteristics of constructive comments and different argumentation styles as described in the literature~\cite{Sukumaran2011, kolhatkar2017b, ting1991, ting1998}. These characteristics were also consistent with open-ended explanations that participants provided in a small pilot (US: 27, India: 22) preceding the main experiment:
\begin{itemize}
    \item More relevant to the original conversation
    \item Uses better logic and facts to support arguments
    \item Takes a better solution-oriented approach
    \item Balances different viewpoints better
    \item Uses more polite and respectful language
    \item Other (please describe):
\end{itemize}

We then repeated the same process with GPT-4. While participants in India and the US reviewed threads that were relevant to their cultural contexts, GPT-4 reviewed threads from both countries.
Table~\ref{tab:phase1-demo} shows the demographic details of the participants. In total, 103 people participated in the study. We removed responses from four participants, who failed the attention check. Participants received $\$1$ for completing the task. We conducted multiple Chi-square tests with Bonferroni corrections to analyze the responses from both participants and GPT-4 to see if their perceptions of constructiveness differ depending on different argumentation styles.

\begin{table}[t]
\caption{Demographic details of participants evaluating constructive comments in Phase 1.}
\label{tab:phase1-demo}
\resizebox{\columnwidth}{!}{
\begin{tabular}{|c|c|c|c|c|}
\hline
\textbf{Demographic (n)} & \textbf{Age (years)} & \textbf{Gender}                                                   & \textbf{Views on same-sex marriage}                                                   & \textbf{Views on Islam}                                                               \\ \hline
US (n=50)               & 43.92 (SD: 11.68)    & \begin{tabular}[c]{@{}c@{}}Female: 48\%\\ Male: 52\%\end{tabular} & \begin{tabular}[c]{@{}c@{}}Against: 29\%\\ Neutral: 8\%\\ Support: 63\%\end{tabular}  & \begin{tabular}[c]{@{}c@{}}Against: 31\%\\ Neutral: 27\%\\ Support: 42\%\end{tabular} \\ \hline
India (n=49)             & 29.08 (SD: 8.75)     & \begin{tabular}[c]{@{}c@{}}Female: 44\%\\ Male: 56\%\end{tabular} & \begin{tabular}[c]{@{}c@{}}Against: 29\%\\ Neutral: 25\%\\ Support: 46\%\end{tabular} & \begin{tabular}[c]{@{}c@{}}Against: 60\%\\ Neutral: 12\%\\ Support: 28\%\end{tabular} \\ \hline
\end{tabular}}
\end{table}

\subsection{Phase 2: Writing Constructive Comments}
We next examined if LLMs could help Indian and American participants write constructive comments on divisive social issues,  
such as homophobia and Islamophobia.

\parabold{Writing Task} We designed a between-subjects experiment where participants were randomly assigned to either a control or a test group. Within each group, participants were first asked about their views on same-sex marriage and Islam (using the same instrument from Phase 1). Then they were shown one randomly selected homophobic and one Islamophobic thread relevant to their cultural contexts. We randomized the order of the threads and asked participants to write a constructive comment for each thread. Since constructive comments tend to be longer~\cite{Sukumaran2011, kolhatkar2017b}, we asked the participants to write a comment that is at least 50 words long. We also disabled copy-paste so that participants could not look up the threads online and submit copied responses. 

\begin{figure*}
    \centering
    \includegraphics[width=\linewidth]{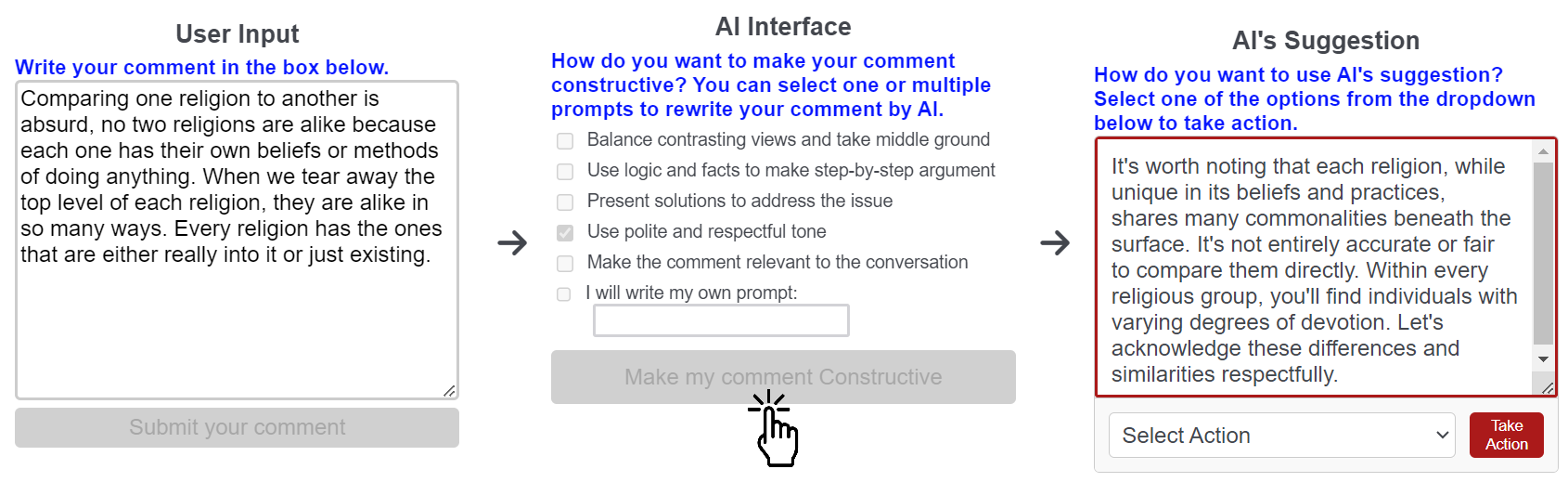}
    \caption{The interface for co-writing constructive comments with AI (LLM) in the test group. Participants first wrote their comment in the \textbf{User Input} box. They could select one or more prompts from the \textbf{AI Interface}. After they selected the prompt and clicked the button ``Make my comment Constructive'', GPT-4 would rewrite participant's comment in real-time---which would appear in the \textbf{AI Suggestion} box. Participants were required to either accept, reject, regenerate, or edit AI's suggestion before submitting their final comment or prompting AI again. They could repeat the process as many times as needed before submitting their final comment. The example shown is based on a comment written by an American participant in response to an Islamophobic thread.}
    \label{fig:Phase-2-writing}
    \Description{User Input: "Comparing one religion to another is absurd, no two religions are alike because each one has their own beliefs or methods of doing anything. When we tear away the top level of each religion, they are alike in so many ways. Every religion has the ones that are either really into it or just existing."
    User selected: "Use polite and respectful tone" to make my comment constructive.
    User received suggestion from AI: "It's worth noting that each religion, while unique in its beliefs and practices, shares many commonalities beneath the surface. It's not entirely accurate or fair to compare them directly. Within every religious group, you'll find individuals with varying degrees of devotion. Let's acknowledge these differences and similarities respectfully."}
\end{figure*}

In the \textbf{control group}, participants were asked to write constructive comments on their own without any external assistance (e.g., the Internet or ChatGPT). 
In the \textbf{test group}, we instructed participants to write an initial draft before requesting help from an LLM (GPT-4) (see Figure~\ref{fig:Phase-2-writing}). Once they wrote a draft, they could select from a list of prompts or provide custom prompt for the LLM to rewrite their comment constructively. The prompts were created to reflect characteristics of constructive comments and logical and dialectical argumentation, as described in prior work~\cite{Sukumaran2011, kolhatkar2017b, ting1991, ting1998}:
\begin{itemize}
    \item Make my comment relevant to the conversation
    \item Use logic and facts to make step-by-step argument
    \item Present solutions to address the issue
    \item Balance contrasting views and take middle ground
    \item Use polite and respectful tone
    \item Other (write your own prompt):
\end{itemize}

We randomized the order of the prompts to avoid primacy and recency bias. To ensure that participants in the test group used LLM while writing constructive comments, they were required to prompt the LLM at least once before completing the writing task. Furthermore, participants had to write at least 20 words before requesting suggestions from the LLM,  ensuring that the resulting comments reflected genuine human-AI (HAI) collaboration. We used GPT-4 to rewrite participant's comment on the given thread in real-time using the following prompt:
\begin{quote}
    \textit{Consider the following Reddit thread:}
    \\
    \textit{<insert thread>}

    \textit{An <Indian, American> participant, who think <insert stance> of <homophobia, Islamophobia> wrote the following comment in response to the above thread.}
    \\
    \textit{<insert participant's comment>}

    \textit{Make the comment constructive using following prompts.}
    \\
    \textit{<insert prompts selected or written by the participant>}
\end{quote}

After receiving the suggestion from the underlying LLM, participants could either accept, reject, regenerate, or edit the comment before they could prompt the LLM again. Once participants were satisfied with the final output, they could submit it to finish the task. 


\parabold{Post-Writing Survey} Prior study shows that the ease of writing with LLMs can reduce users' sense of ownership over their writing~\cite{Kim-2023}. Therefore, after finishing the writing task, participants in both groups were asked to reflect how much ownership and satisfaction they felt with the final comments on a 5-point Likert scale. They also rated the difficulty of writing constructive comments. In addition, participants in the test group were shown one randomly selected LLM-generated suggestion that they either accepted, rejected, regenerated, or edited. They were then asked to explain the reasoning behind their choice. 

\parabold{Participant Recruitment} To complete the writing task, we recruited 52 Indian and 52 American 
crowd workers from Prolific, who did not take part in Phase 1. Table~\ref{tab:phase2-demo} shows the demographic details of the participants.
We compensated participants with $\$1.70$ for completing the writing task. Overall, we collected 104 human-written comments from the control group and 102 human-AI (HAI) written comments from the test group.

\begin{table}[t]
\caption{Demographic details of participants, who completed the writing task in Phase 2.}
\label{tab:phase2-demo}
\resizebox{\columnwidth}{!}{
\begin{tabular}{|c|c|c|c|c|}
\hline
\textbf{Demographic (n)} & \textbf{Age (years)} & \textbf{Gender}                                                                      & \textbf{Views on same-sex marriage}                                                   & \textbf{Views on Islam}                                                               \\ \hline
US (n=52)               & 38.79 (SD: 13.19)    & \begin{tabular}[c]{@{}c@{}}Female: 60\%\\ Male: 38\%\\ Transgender: 2\%\end{tabular} & \begin{tabular}[c]{@{}c@{}}Against: 15\%\\ Neutral: 10\%\\ Support: 75\%\end{tabular} & \begin{tabular}[c]{@{}c@{}}Against: 23\%\\ Neutral: 29\%\\ Support: 48\%\end{tabular} \\ \hline
India (n=51)             & 28.63 (SD: 7.88)     & \begin{tabular}[c]{@{}c@{}}Female: 27\%\\ Male: 73\%\end{tabular}                    & \begin{tabular}[c]{@{}c@{}}Against: 43\%\\ Neutral: 22\%\\ Support: 35\%\end{tabular} & \begin{tabular}[c]{@{}c@{}}Against: 45\%\\ Neutral: 8\%\\ Support: 47\%\end{tabular}  \\ \hline
\end{tabular}}
\end{table}

\parabold{Human Evaluation of Constructiveness} 
We next conducted a survey to examine which type of comment people perceived as more constructive: those written by humans (control group), human-AI (HAI) collaboration (test group), or solely generated by LLM (Phase  1).  We recruited 82 Indian and 82 American crowd workers from Prolific, who had not participated in both Phase  1 and the writing task in Phase  2. Each participant was shown a randomly chosen homophobic or Islamophobic thread related to their cultural context. Then, they were asked to review four pairs of randomly chosen comments that were written for the same thread from the same stance. Each pair could either include (HAI vs. Human), (Human vs. AI), or (HAI vs. AI) comments. 

Each participant saw each comment only in one pair and we randomized the order of comments within each pair. We asked participants to select which comment they perceived as more constructive within each pair. We ensured that each comment was reviewed by at least three participants. 
In total, we received 727 human evaluations from 157 participants after discarding responses from 7 participants, who failed the attention check. Participants were compensated with $\$1.00$ for completing the task.

\parabold{Quantitative Analysis} 
Next, we analyzed the key linguistic features of constructiveness (see Table~\ref{tab:constructive-comment-feature}) in comments written by humans (control) and those via human-AI collaboration (test). We followed the same procedure as in Phase 1 where we calculated these features in LLM-generated comments. For the test group, 
we also analyzed how incorporating LLM's suggestions impacted the quality of comments by comparing participants' initial drafts with their final submissions. Specifically, we compared the linguistic features of constructiveness, sentiment, and toxicity between initial and final comments. To analyze sentiment, we used VADER sentiment analysis tool as it is attuned to sentiments expressed in social media~\cite{hutto2014}. To detect toxicity, we used Google's Perspective API. We also calculated cosine similarity, semantic similarity, and BERTScore~\cite{Zhang-2019} between these pairs of comments. For cosine similarity, we used CountVectorizer from scikit-learn\footnote{https://scikit-learn.org/1.5/modules/generated/sklearn.feature{\textunderscore}extraction.text.CountVectorizer.html}. For semantic similarity, we used all-MiniLM-L6-v2 sentence transformer\footnote{https://huggingface.co/sentence-transformers/all-MiniLM-L6-v2} from Hugging Face. For BERTScore, we used contextual embeddings in BERT model\footnote{https://huggingface.co/docs/transformers/en/model{\textunderscore}doc/bert} from Hugging Face.

\parabold{Ethical Considerations} Since the study involved engaging with sensitive and potentially hateful content, we took several measures to conduct our research ethically and responsibly. When posting the tasks on Prolific, we included a content warning for sensitive topics, allowing participants to make an informed decision before opting into the task. We also used Prolific's Harmful Content Prescreener so that the tasks are only shown to participants who had self-reported being comfortable with sensitive content. At the outset of the study, we explicitly informed participants that they would be reviewing threads related to homophobia and Islamophobia. We emphasized the potential emotional impact of engaging with such content and assured them that they could withdraw at any time if they felt uncomfortable. This was communicated before obtaining informed consent or collecting any data. Within the study, after gathering participants' views on same-sex marriage and Islam, we provided a trigger warning before displaying the threads. This was to remind participants that they could opt out at any point if they chose to do so. Moreover, the study was designed in a way that gave participants the freedom to write based on their personal stances. We encouraged the participants to share their honest opinions and communicated that the study does not aim to condone or criticize anyone's views. Additionally, we provided country-specific resources to help participants manage emotional distress and shared contact details for any study-related concerns.

\section{Findings}
We first describe results from Phase 1 to answer how perceptions of constructiveness vary depending on logical vs. dialectical argumentation style (Section \ref{rq1findings}). We then present our analyses of the comments participants wrote in Phase 2 to examine if LLMs could help people from different cultures write constructive comments on divisive social issues (Section \ref{rq2findings}). 

\subsection{RQ1: Do Perceptions of Constructiveness Differ Based on Argumentation Styles?}
\label{rq1findings}
\begin{figure*}
    \centering
    \includegraphics[width=\linewidth]{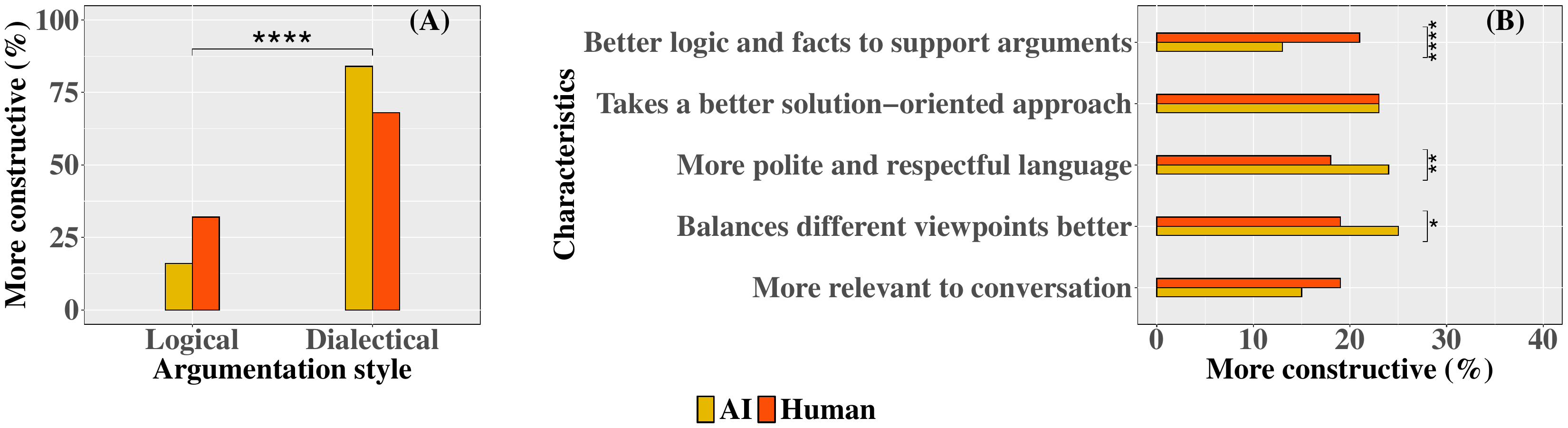}
    \caption{(A) Perceptions of constructive comments between humans and LLM based on argumentation style. (B) Perceived characteristics of constructive comments reported by both humans and LLM. Statistically significant differences are reported at $p<0.00001$ (****), $p<0.0001$ (***), $p<0.001$ (**), and $p<0.01$ (*) [adjusted P-values after Bonferroni correction].}
    \label{fig:human-ai-phase1}
    \Description{(A) A bar chart showing LLM considered dialectical comments as more constructive than logical comments 84\% time, whereas humans considered dialectical comments as more constructive 68\% time. (B) A bar chart showing the distribution of perceived characteristics of constructive comments by humans and LLM. }
\end{figure*}

\parabold{Preference Between Logical and Dialectical Comments} In Phase 1, participants reviewed 396 pairs of logical and dialectical comments, and GPT-4 reviewed 454 such pairs. Binomial tests revealed that participants as well as GPT-4 found dialectical comments to be more constructive than logical comments, which significantly differed ($p<0.000001$) from the hypothesized proportion (50\%). GPT-4 selected dialectical comments as more constructive than logical ones in 84\% of cases, whereas participants selected dialectical comments as more constructive in 68\% of cases (see Figure~\ref{fig:human-ai-phase1}A). A chi-square test with Yates' continuity correction revealed a significant difference between GPT-4 and participants' preference for logical vs. dialectical comments as constructive  ($\chi^{2}(1, N=850)=28.52, p<0.00001, \phi=0.18$, odds ratio=2.46), revealing a small effect size.

GPT-4's preference for dialectical comments as constructive also differed significantly from that of both American and Indian participants, who viewed dialectical comments as more constructive in 73\% and 65\% of cases, respectively.  
American participants' preference for dialectical arguments over logical comments contradicts observation from previous research, which suggests that people from individualistic cultures (e.g., the US) prefer logical reasoning during argumentation~\cite{norenzayan2002cultural}. A binomial test revealed that American participants' preference for dialectical comments significantly differed ($p<0.000001$) from the expected distribution (40\% as observed in~\citet{peng1997naive}).

\parabold{Perceived Characteristics of Constructiveness} When we asked participants and GPT-4 why their chosen comment was more constructive than the other, on average they selected three characteristics from the options given to them (See Figure~\ref{fig:human-ai-phase1}B). While assessing constructive comments, participants prioritized the use of logic and facts (21\%) and presence of solutions (23\%) more than other factors. Both American and Indian participants reported similar reasons (see Table~\ref{tab:char-in-US} in Appendix), suggesting a shared understanding of what counts as constructive across different cultures. In contrast, GPT-4 focused more on how well the comments balanced different viewpoints (25\%) and maintained a polite tone (24\%). A chi-square test with Yates' continuity correction revealed a significant difference in the perceived characteristics of constructive comments between participants and GPT-4 ($\chi^{2}(4, N=2824)=45.68, p<0.00001, \phi=0.13$), with a small effect size. Post-hoc tests with Bonferroni corrections showed that GPT-4 associated the use of polite and respectful language (24\% of cases) and balancing different viewpoints (25\% of cases) with constructiveness significantly more than the participants (18\% of cases for both characteristics). These characteristics are associated with dialectical argumentation and constructive discourse, which align with GPT-4's greater preference for dialectical comments as constructive. 

Together, these findings indicate potential misalignment in how humans and LLMs characterize constructiveness. While both perceived comments with dialectical arguments as more constructive than those with logical arguments, GPT-4 was 2.46 times more likely than humans to rate dialectical comments as more constructive. GPT-4 also had significantly different perceptions of what makes an online comment constructive. These differences could impact people's engagement with LLM while writing constructive comments on divisive social issues, which we examine next.

\subsection{RQ2: Can LLM Help People Write Constructive Comments?}
\label{rq2findings}
In Phase 2, we collected 104 human-written comments (control group) and 102 HAI-written comments (test group) on homophobic and Islamophobic threads. Additionally, we had 64 LLM-generated comments from Phase 1 for these threads. We now present our analysis of these comments.

\subsubsection{Who Writes Constructive Comments Better?} 
To assess which type of comment---human-written (Human), HAI-written (HAI), or LLM-generated (AI)---was perceived as more constructive, we conducted human evaluations and analyzed linguistic features associated with constructiveness across the three sets of comments.

\parabold{Human Evaluation of Constructiveness} A total of 157 participants evaluated 727 pairs of comments, which included 401 pairs of HAI vs. Human comments, 141 pairs of Human vs. AI comments, and 185 pairs of HAI vs. AI comments. Multiple Chi-square tests with Bonferroni corrections revealed significant differences in people's perceptions of constructiveness across different comment pairs (see Table~\ref{tab:chi-sq-constructive}). 

\begin{figure}[t]
    \centering
    \includegraphics[width=\linewidth]{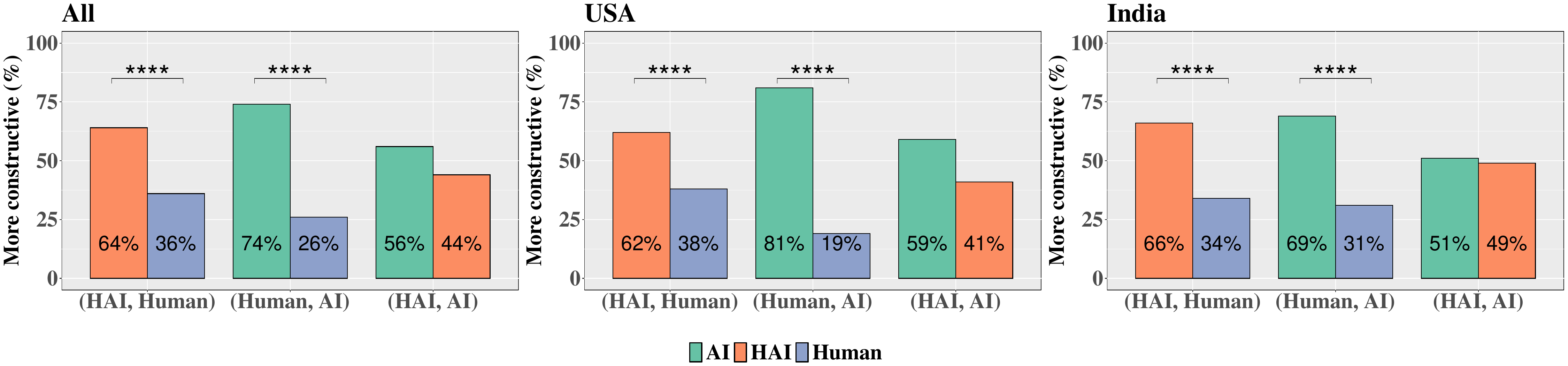}
    \caption{People's perceptions of constructiveness across (Human vs. AI), (HAI vs. Human), and (HAI vs. AI) comment pairs. Statistically significant differences are reported at $p<0.000005$ (****), $p<0.00005$ (***), $p<0.0005$ (**), and $p<0.005$ (*) [adjusted P-values after Bonferroni correction]. }
    \label{fig:human-eval-constructive}
    \Description{Grouped bar charts showing People perceived LLM-generated comments as more constructive than the human-written comments in 74\% of pairs. Indian participants perceived LLM-generated comments as more constructive than the human-written comments in 69\% of pairs. American participants perceived LLM-generated comments as more constructive than the human-written comments in 81\% of pairs. Participants perceived HAI-written comments as more constructive than the human-written comments in 64\% of pairs. Indian participants perceived HAI-written comments as more constructive than the human-written comments in 66\% of pairs. American participants perceived HAI-written comments as more constructive than the human-written comments in 62\% of pairs.}
\end{figure}

Participants who assessed Human vs. AI comments found LLM-generated comments significantly more constructive than human-written comments in the majority of cases (74\%) (see Figure~\ref{fig:human-eval-constructive}), with a medium effect size ($\chi^{2}(1, N=282)=65.59, p<0.000005, \phi=0.48$, odds ratio=8.51). Participants were 8.51 times more likely to choose LLM-generated comments as constructive than human-written comments. 

Both Indian and American participants perceived LLM-generated comments as significantly more constructive than human-written comments in 69\% and 81\% of cases, respectively (see Figure~\ref{fig:human-eval-constructive}).

\begin{table}[t]
\caption{Multiple Chi-square tests with Bonferroni corrections comparing people's perceptions of constructiveness across HAI vs. Human, Human vs. AI, and HAI vs. AI comments.}
\label{tab:chi-sq-constructive}
\resizebox{\columnwidth}{!}{
\begin{tabular}{|l|l|l|l|}
\hline
\multicolumn{1}{|c|}{\textbf{Comment pairs}}                                                                                               & \multicolumn{1}{c|}{\textbf{Both demographic}}                                                & \multicolumn{1}{c|}{\textbf{India}}                                                              & \multicolumn{1}{c|}{\textbf{US}}                                                                \\ \hline
\textbf{HAI vs. AI}                                                                                                                         & \multicolumn{1}{c|}{-}                                                                        & \multicolumn{1}{c|}{-}                                                                           & \multicolumn{1}{c|}{-}                                                                           \\ \hline
\begin{tabular}[c]{@{}l@{}}\textbf{Human vs. AI}:\\ LLM-generated comments perceived as more \\ constructive than human-written comments\end{tabular}   & \begin{tabular}[c]{@{}l@{}}$\chi^{2}(1, N=282)=65.59, $\\ $p<0.000005, \phi=0.48$\end{tabular} & \begin{tabular}[c]{@{}l@{}}$\chi^{2}(1, N=148) = 19.70, $\\ $p<0.000005, \phi=0.36$\end{tabular}  & \begin{tabular}[c]{@{}l@{}}$\chi^{2}(1, N=134)=47.76,$ \\ $p<0.000005, \phi=0.60$\end{tabular}    \\ \hline
\begin{tabular}[c]{@{}l@{}}\textbf{HAI vs. Human}:\\ HAI-written comments perceived as more \\ constructive than human-written comments\end{tabular} & \begin{tabular}[c]{@{}l@{}}$\chi^{2}(1, N=802)=62.56, $\\ $p<0.000005, \phi=0.28$\end{tabular} & \begin{tabular}[c]{@{}l@{}}$\chi^{2}(1, N =205) = 39.96, $\\ $p<0.000005, \phi=0.31$\end{tabular} & \begin{tabular}[c]{@{}l@{}}$\chi^{2}(1, N =392) = 22.54,$ \\ $p<0.000005, \phi=0.24$\end{tabular} \\ \hline
\end{tabular}

}
\end{table}

Similarly, participants, who reviewed HAI vs. Human comment pairs found HAI-written comments significantly more constructive than human-written comments in 64\% of cases, with small effect size ($\chi^{2}(1, N=802)=62.56, p<0.000005, \phi=0.28$, odds ratio=3.19). Participants were 3.19 times more likely to prefer HAI-written comments as constructive compared to human-written comments. 

Both Indian and American participants significantly preferred HAI-written comments to those solely written by humans in 66\% and 62\% of cases, respectively (see Figure~\ref{fig:human-eval-constructive}). 

On the other hand, those who reviewed HAI vs. AI comment pairs considered LLM-generated comments as constructive only in 55\% of cases. We did not notice any significant difference between participants' preferences for LLM-generated and HAI-written comments. 

In sum, these findings suggest that comments that are either fully generated by LLM or co-written by human and LLM were perceived as more constructive than those written by humans alone. 

\begin{figure}[t]
    \centering
     \begin{subfigure}[b]{\textwidth}
         \centering
         \includegraphics[width=\textwidth]{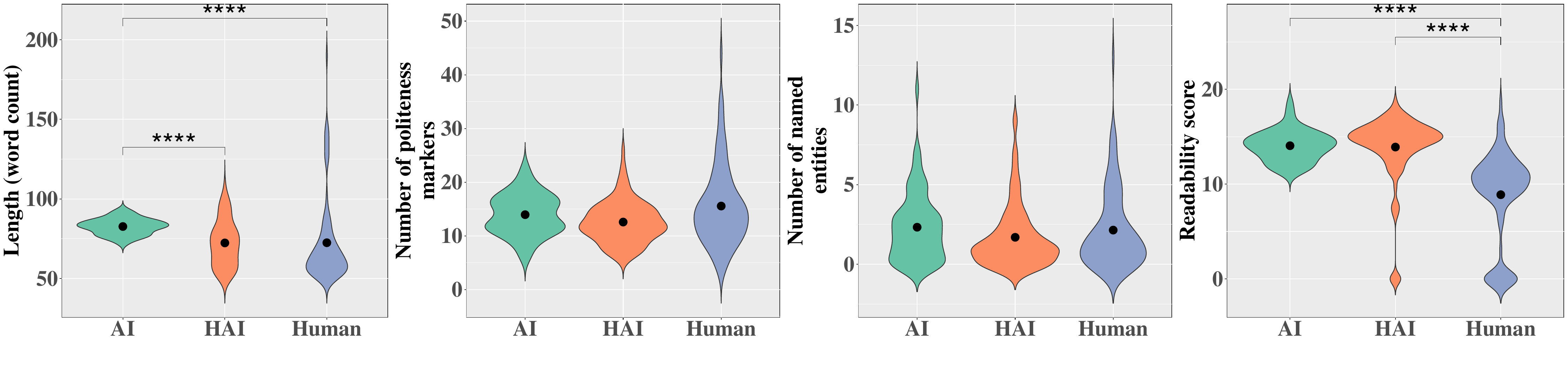}
     \end{subfigure}
     \\
     \begin{subfigure}[b]{\textwidth}
         \centering
         \includegraphics[width=\textwidth]{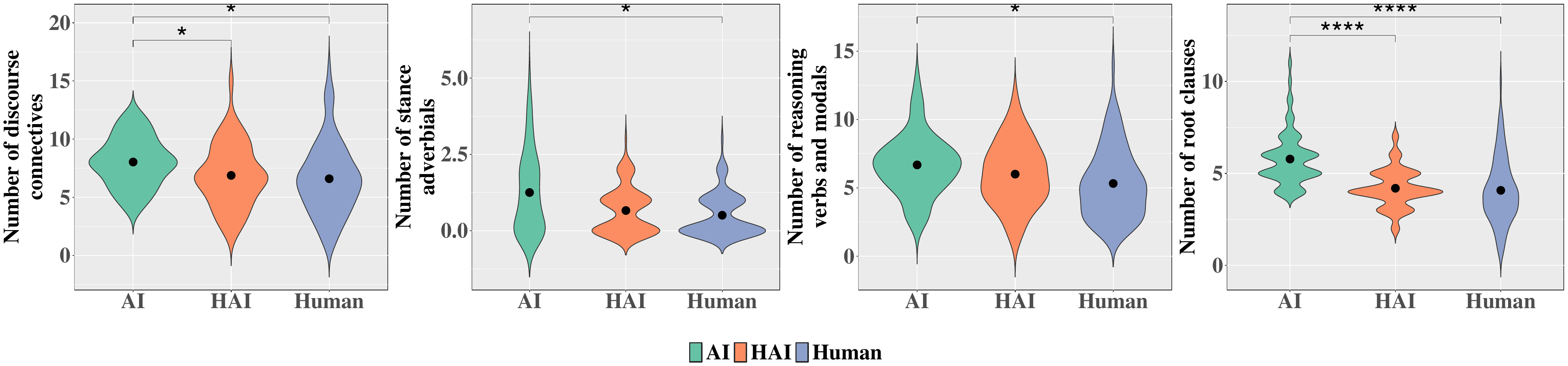}
     \end{subfigure}
    \caption{Features of constructiveness across LLM-generated, human-written, and HAI-written comments. The black dots represent the average value. Statistically significant differences are reported at $p<0.000001$ (****), $p<0.00001$ (***), $p<0.0001$ (**), and $p<0.001$ (*) [adjusted P-value after Bonferroni correction].}
    \label{fig:constructive-features}
    \Description{Eight grouped violin plots featuring how length, politeness, named entities, readability score, discourse connectives, stance adverbials, reasoning verbs and modals, and root clauses vary among LLM-generated, human-written, and HAI-written comments.}
\end{figure}

\parabold{Linguistic Features of Constructiveness} To examine how constructiveness varies among LLM-generated, human-written, and HAI-written comments, we computed values for different linguistic features and conducted multiple pairwise Mann-Whitney tests with Bonferroni corrections (see Table~\ref{tab:constructive-feature} in Appendix). We found that LLM-generated comments were significantly more constructive, i.e., they were longer and included more argumentative features, such as discourse connectives, stance adverbials, reasoning verbs and modals, and root clauses compared to human-written comments (see Figure~\ref{fig:constructive-features}). LLM-generated comments were also longer, contained more discourse connectives and root clauses than HAI-written comments. The effect size ranged from small to large. 

 We wanted to check if the greater presence of argumentative features in LLM-generated comments was due to greater length. We conducted a multivariate analysis of covariance (MANCOVA) to examine whether comment type had a significant effect on different argumentative features, while controlling for the comment length. The results showed a significant multivariate effect of comment type (Pillai's Trace = 0.34, $F(8, 506) = 12.88, p<0.000001$)--indicating that 34\% of the multivariate variance in the set of argumentative features is associated with comment type after adjusting for length. See Table~\ref{tab:ancova-test} in Appendix for detailed statistics.


Additionally, we noted that both LLM-generated and HAI-written comments had significantly higher readability score (as measured by SMOG index) than human-written comments. This suggests that comments that are either fully generated by LLM or collaboratively written with LLM have better wording and sentence structure than human-written comments. 

\parabold{Time Taken to Write Constructive Comments} 
Participants who wrote comments with LLM (test group) took less time (average: 5.17 minutes) to write constructive comments than those who wrote everything on their own in control group (average: 6.71 minutes). A Mann-Whitney's U test revealed a significant but small effect of LLM intervention on the time taken to write constructive comments ($W=4071, Z=-2.88, p<0.01, r=0.2$). While this aligns with previous findings showing that LLMs help people write faster~\cite{Kim-2023}, our results indicate that LLM assistance not only speeds up writing but also enhances constructiveness.  


\subsubsection{How does Co-Writing with LLM Change the Quality of Comments?}\label{sec:cowrite-change} 
In the test group, participants requested assistance from LLM 138 times to write 102 constructive comments using 300 prompts in total (average: 2.88, SD: 2.36). Almost all participants chose the given prompts (see Table \ref{tab:prompts-used}). They prioritized relevance, use of logic, and politeness to make their comments constructive. Only three participants wrote custom prompts, such as: \textit{`Make them sound dumb for being so bigoted', `Be passive aggressive', `Make it less rude', and `Make the comment kind.'} 

Although all participants in the test group were required to prompt the LLM, 87 of them incorporated LLM's suggestions in their writing. We compared these participants' initial drafts to the final HAI-written comments submitted by them.

\begin{table}[t]
\caption{Different prompts used by participants to make their comments constructive in the test group}
\label{tab:prompts-used}
\begin{tabular}{|l|c|}
\hline
\multicolumn{1}{|c|}{\textbf{Prompts}}            & \textbf{Used (\%)} \\ \hline
Make the comment relevant to the conversation     & 23                 \\ \hline
Use logic and facts to make step-by-step argument & 23                 \\ \hline
Present solutions to address the issue            & 18                 \\ \hline
Balance contrasting views and take middle ground  & 12                 \\ \hline
Use polite and respectful tone                    & 23                 \\ \hline
Custom prompt                                     &  1                 \\ \hline      
\end{tabular}
\end{table}

\parabold{Content Similarity} 
We first examined to what extent LLM changed participants' initial comments to the ones they submitted finally. 
For this, we calculated semantic similarity, BERTScore~\cite{Zhang-2019}, and cosine similarity between each participant's initial draft (before requesting help from LLM) and the final comment they submitted (rewritten by LLM). The values of these measures range from 0 to 1, with higher values indicating greater similarity. 

\begin{figure}[t]
    \centering
     \begin{subfigure}[b]{\textwidth}
         \centering
         \includegraphics[width=\textwidth]{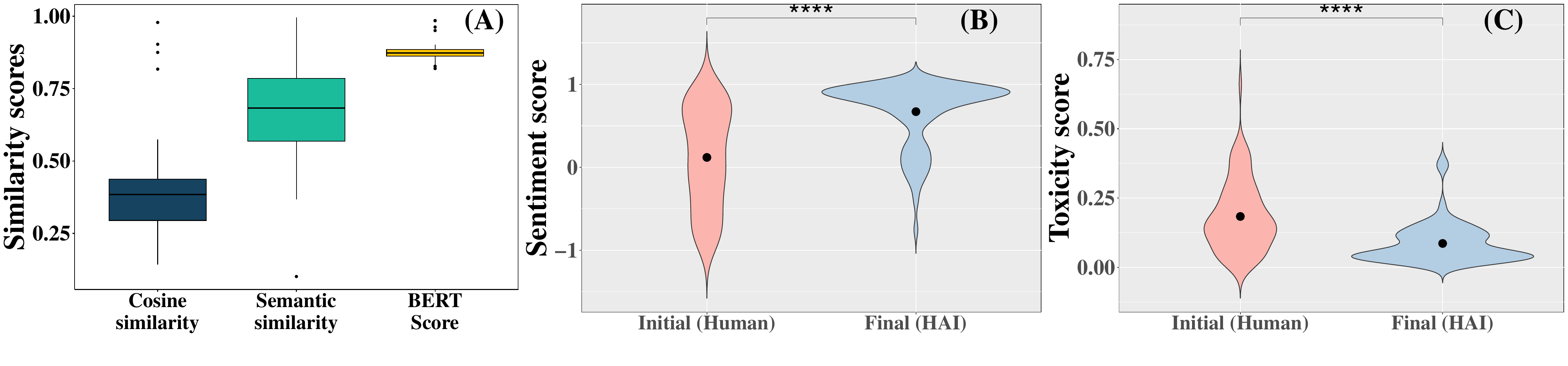}
     \end{subfigure}
     \\
     \begin{subfigure}[b]{\textwidth}
         \centering
         \includegraphics[width=\textwidth]{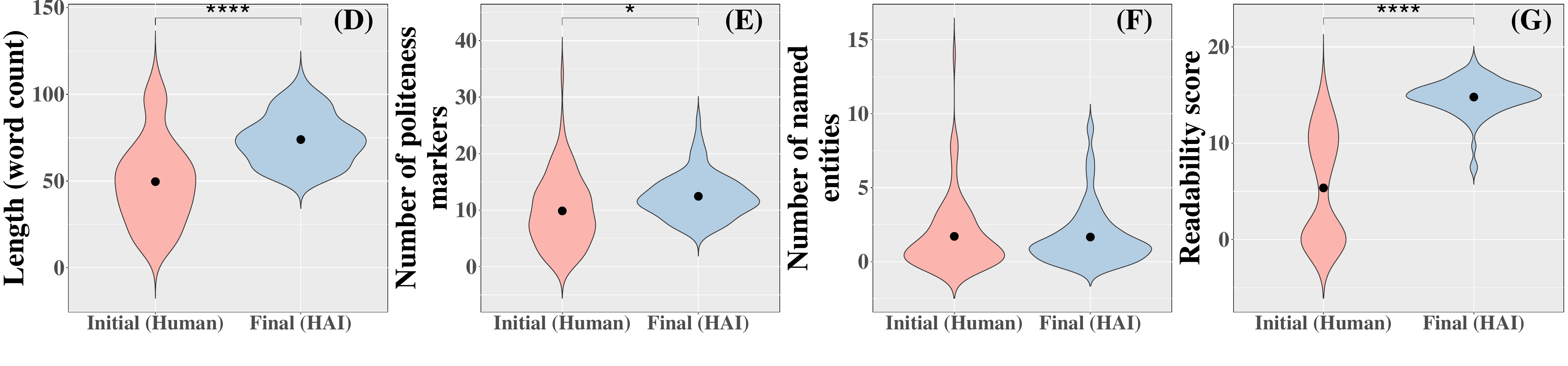}
     \end{subfigure}
     \\ 
     \begin{subfigure}[b]{\textwidth}
         \centering
         \includegraphics[width=\textwidth]{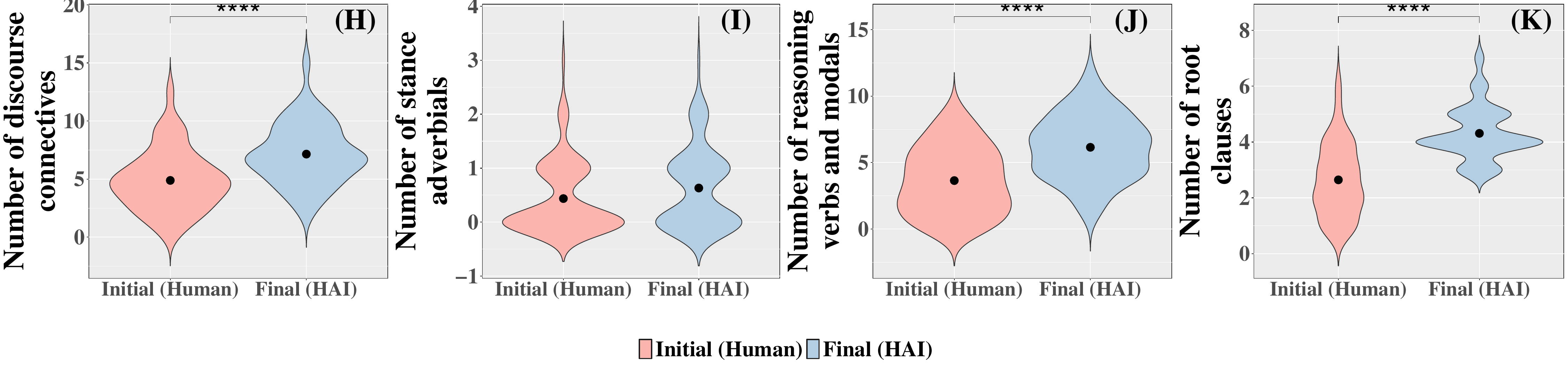}
     \end{subfigure}     
    \caption{Comparison between different characteristics of comments that participants initially wrote themselves in the test group and the comment they submitted, where they used LLM's suggestion. The black dots represent the average value. Statistically significant differences are reported at $p<0.000001$ (****), $p<0.00001$ (***), $p<0.0001$ (**), and $p<0.001$ (*) [adjusted P-value after Bonferroni correction].}
    \label{fig:initial-final-comparison}
    \Description{Eight grouped violin plots featuring how length, politeness, named entities, readability score, discourse connectives, stance adverbials, reasoning verbs and modals, and root clauses vary among LLM-generated, human-written, and HAI-written comments.}
\end{figure}

We found that the average semantic similarity between these comments was 0.67, indicating LLM could potentially retain the meanings of initial comments that participants wrote themselves (see Figure~\ref{fig:initial-final-comparison}A). Similarly, the average high BERTScore of 0.87 indicate that the LLM-rewritten versions closely aligned with the meanings in initial comments. However, the average cosine similarity between these comments was 0.38, indicating potential differences between the wording of these comments. For example, our analysis found that LLM-rewritten version of participants' comments were significantly more positive and less toxic (average sentiment score: 0.67, toxicity score: 0.08) compared to the initial comments people wrote (average sentiment score: 0.12, toxicity score: 0.18) (see Figure~\ref{fig:initial-final-comparison}B--C).

\parabold{Linguistic Features of Constructiveness} We also analyzed how LLM affected the linguistic features of constructiveness while rewriting participants' comments (see Table~\ref{tab:cultural-arg-result} in Appendix). We found that LLM-rewritten versions were significantly longer (average length: 73 words) than the initial comments (average length: 49 words) and the effect size was large (see Figure~\ref{fig:initial-final-comparison}D). Additionally, LLM integrated significantly more argumentative features, such as discourse connectives, reasoning verbs and modals, and root clauses while rewriting participants' comments, with medium to large effect size (see Figure~\ref{fig:initial-final-comparison}H--K).  We wanted to check if the greater presence of argumentative features in final HAI comments was due to greater length. A multivariate analysis of covariance (MANCOVA) showed a significant effect of LLM's intervention (Pillai's Trace = 0.60, $F(4, 168) = 62.93, p<0.000001$)--indicating that 60\% of the multivariate variance in argumentative features is associated with LLM's intervention after adjusting for length. See Table~\ref{tab:ancova-test-hai} in Appendix for detailed statistics.

We also found that LLM made participants' comments significantly more polite (average score: 12.44) than their initial drafts (average score: 9.86), with medium effect size (see Figure~\ref{fig:initial-final-comparison}E). LLM-rewritten comments also had significantly higher readability score (average: 14.79) 
than the initial drafts participants wrote (average score: 5.85), exhibiting large effect size (see Figure~\ref{fig:initial-final-comparison}G). These differences suggest that LLM could present participants' responses more constructively than they could do on their own.

\subsubsection{Does Co-Writing Comments with LLM Lead to Homogeneity?} Recent studies have found that LLM-assisted writing tools homogenize outputs, i.e., different users relying on the same LLM may produce more similar writing than they would without such assistance \cite{Kreminski-2024}. To examine whether co-writing constructive comments with LLM leads to homogenity, we analyzed the cosine similarity, semantic similarity, and ROUGE-L~\cite{padmakumar2024} scores among HAI-written comments written on the same thread by different participants who share same stance. We repeated the same analysis for human-written comments in the control group to establish our baseline.

We found that the cosine similarity and semantic similarity of HAI-written comments in the test group were comparable to the similarity scores of human-written comments in the control group (see Table~\ref{tab:homogeneity-metric}). Multiple Mann-Whitney U tests with Bonferroni corrections did not show significant difference in these metrics between control and test groups. This indicates that even when participants used LLM assistance to write constructive comments on the same thread from the same stance, the overlap in both wording (cosine similarity) and meaning (semantic similarity) among their comments was comparable to that of comments written without LLM assistance.

\begin{table}[t]
\caption{Results from Mann-Whitney U tests with Bonferroni corrections for different metrics to assess homogeneity in HAI-written comments.}
\label{tab:homogeneity-metric}
\resizebox{\columnwidth}{!}{
\begin{tabular}{|l|l|c|c|}
\hline
\multicolumn{1}{|c|}{\textbf{Metric}} & \multicolumn{1}{c|}{\textbf{Statistics}}       & \textbf{\begin{tabular}[c]{@{}c@{}}Human-written \\ comments (control)\end{tabular}} & \textbf{\begin{tabular}[c]{@{}c@{}}HAI-written \\ comments (test)\end{tabular}} \\ \hline
Cosine similarity                     & \multicolumn{1}{c|}{-}                         & 0.33                                                                                 & 0.30                                                                                  \\ \hline
Semantic similarity                   & \multicolumn{1}{c|}{-}                         & 0.41                                                                                 & 0.42                                                                                  \\ \hline
Rogue-L                               & U=42029, Z=-3.73, p\textless{}0.001, r=0.16    & 0.18                                                                                 & 0.16                                                                                  \\ \hline
Distinct-2                            & U=2153.5, Z=6.88, p\textless{}0.000001, r=0.50 & 0.97                                                                                 & 0.99                                                                                  \\ \hline
Type token ratio                      & U=1247.5, Z=8.66, p\textless{}0.000001, r=0.63 & 0.79                                                                                 & 0.88                                                                                  \\ \hline
\end{tabular}}
\end{table}

The ROUGE-L score (i.e., longest common subsequence) among human-written comments (average: 0.18)  was significantly higher than that of HAI-written comments (average: 0.16). This suggests that the comments participants wrote independently in the control group had greater overlap than the comments participants co-wrote using LLM in the test group. 
In fact, both distinct-2 score and type token ratio (TTR) were significantly higher for the HAI-written comments, signaling greater diversity of vocabulary in these comments. Together, these findings suggest that using LLM to write constructive comments on socially divisive issues may not necessarily lead to homogeneous discourse online. This is critical because overwhelming agreement and homogeneity in online threads discourage people from engaging with that thread~\cite{Grevet2014}.

\subsubsection{How did People Interact with LLM while Co-Writing Comments?} 
Next, we examined participants' responses to the changes LLM made to their initial drafts in the test group. We quantitatively analyzed the interaction logs and performed qualitative analyses on the open-ended reasons participants provided for accepting, editing, rejecting, or regenerating the LLM's suggestions.

\parabold{Acceptance} In most cases (62\%), participants accepted LLM-generated suggestions. They found the suggestions captured their main points clearly ($n=18$). Participants ($n=9$) reported that LLM-generated versions were more persuasive, well-worded (higher readability score as observed in Figure~\ref{fig:initial-final-comparison}G), and \textit{``presented cohesive arguments with actionable insights.''} Some participants (n=4) appreciated that AI saved their time and effort, noting that it would take them significantly longer to \textit{``refine their thoughts''} if they were to write constructively on their own. Others ($n=4$) valued the polite and impartial tone in comments re-written by LLM. An American participant wrote:
\begin{quote}
    \textit{I like AI assistance because it removes my own potential bias in these settings. I think being dismissive and defensive is never forward thinking. It is nice to remove some emotional substance when it is not needed.}
\end{quote}

\parabold{Editing} In 9\% of cases, participants either deleted some parts of the LLM-generated content, changed wording (e.g., replaced `same-sex' with `gay', `concur' with `agree'), or added new content. They explained that the LLM re-written version was either \textit{``too moderate''} or misrepresented their views. \color{change}Upon manual review, we identified 12 instances (12.7\%) where LLM changed people's original opinion when rewriting their comments constructively---especially when the views were in a spectrum. \color{black}An Indian participant, who was against same-sex marriage, shared: 
\begin{quote}
\textit{I wrote about respecting LGBTQ communities and protecting their rights. But I strongly feel that legalizing LGBTQ marriages will imbalance both the culture and the nature. AI misunderstood my comment and wrote in favor of legalizing such marriages.}
\end{quote}

On average, people edited LLM's suggestions in a way that made the comments significantly less positive (mean sentiment: 0.21) and more toxic (mean toxicity: 0.11) than the original suggestion (mean sentiment: 0.78, mean toxicity: 0.08). Paired t-tests with Bonferroni corrections revealed a significant effect of editing on the sentiment ($t(9)=3.83, p<0.01$) and toxicity ($t(9)=-2.82, p<0.01$) of these comments. This indicates that LLMs might not be able to fully capture the diverse spectrum and nuances in a person's stance on divisive issues.

\parabold{Rejection and Regeneration} 
Participants 
regenerated and rejected LLM's suggestions in 11\% and 13\% of cases, respectively. In the post-writing survey, participants reported that they regenerated suggestions because they were either verbose, \textit{``politically correct''}, or used \textit{``HR-sounding language.''} Those, who rejected the suggestions, either did not like the \textit{``formal and essay-like''} and \textit{``non-human and AI-sounding''} language. In several cases, participants disagreed with the framing in LLM-generated suggestions. An Indian participant, who thought Islam incites extremism among its followers, reported:

\begin{quote}
    \textit{I talked about how Muslims are very `hardcore' compared to followers of other religion. But AI framed it as `deep-seated religious commitment' to give it a positive spin, which I didn't like. I also wrote about having uniform civil code to remove the exemption that Muslims enjoy regarding multiple marriages. But AI wrote about codifying such exemptions under uniform civil code, which I don't agree with.}
\end{quote}

\subsubsection{Experience of Writing Constructive Comments} 
Table~\ref{tab:experience-writing} shows that participants in the test group, who used LLM while writing constructive comments, were more satisfied with the comments (82\%) than those who wrote on their own in the control group (77\%). As expected, more participants in the control group (79\%) reported feeling ownership over their comments than those in the test group (76\%). However, these differences in satisfaction and ownership were not statistically significant.
On average, more participants in the control group (31\%) reported that writing constructive comments was difficult on their own than those who received LLM's suggestions (16\%). Despite this gap, the difference in perceived difficulty was not statistically significant either.
This indicates that LLM has the potential to help people write constructive comments on polarizing topics without significantly impacting the value people find while expressing their opinions.

\begin{table}[t]
\caption{Participants' experiences of writing constructive comments}
\label{tab:experience-writing}
\resizebox{\columnwidth}{!}{
\begin{tabular}{l|c|c|c|}
\cline{2-4}
                                                        & \multicolumn{1}{l|}{\textbf{Satisfied (\%)}} & \multicolumn{1}{l|}{\textbf{Feeling of ownership (\%)}} & \multicolumn{1}{l|}{\textbf{Perceived difficulty (\%)}} \\ \hline
\multicolumn{1}{|l|}{Human-written comments (Control)}  & 77                                           & 79                                                      & 31                                                      \\ \hline
\multicolumn{1}{|l|}{HAI-written comments (Test)} & 82                                           & 76                                                      & 16                                                      \\ \hline
\end{tabular}}
\end{table}

\subsubsection{Do Cultural Differences Affect The Writing of Constructive Comments?} We compared the comments written by Indian and American participants, both within control and test groups. We did not find any significant difference in the linguistic features of constructiveness between the comments written by Indian and American participants (see Table~\ref{tab:char-control-test} in Appendix). In the test group, both Indian and American participants also used the LLM prompts in similar ways to make their comments constructive (see Table~\ref{tab:prompts-in-US}). The three most frequently used prompts across both groups were: \textit{``making the comment relevant to the conversation''}, \textit{``using logical argument''}, and \textit{``polite and respectful tone.''} These findings suggest that, despite known cultural differences in argumentation styles, people from both India and the US
may converge in their approaches to writing constructive comments on divisive social issues.

\begin{table}[t]
\caption{Prompts used by Indian and American Participants in the test group}
\label{tab:prompts-in-US}
\resizebox{0.7\columnwidth}{!}{
\begin{tabular}{|l|c|c|}
\hline
\textbf{Prompts}                                  & \multicolumn{1}{l|}{\textbf{India (\%)}} & \multicolumn{1}{l|}{\textbf{US (\%)}} \\ \hline
Make the comment relevant to the conversation     & 27                                       & 21                                     \\ \hline
Use logic and facts to make step-by-step argument & 21                                       & 25                                     \\ \hline
Present solutions to address the issue            & 19                                       & 17                                     \\ \hline
Balance contrasting views and take middle ground  & 10                                       & 14                                     \\ \hline
Use polite and respectful tone                    & 23                                       & 24                                     \\ \hline
\end{tabular}}
\end{table}

However, we did observe some cultural differences in the writing experience. In the control group, 
American participants took less time (5.2 minutes) to write constructive comments than Indian participants (6.7 minutes). A Mann-Whitney U test revealed a statistically significant, though small, effect of culture on writing time  ($U=6742, Z=3.36, p<0.001, r=0.23$). This may be due to English not being the native language for most Indian participants.

In the test group, Indian participants reported significantly higher levels of satisfaction and ownership over their final comments than American participants (see Table~\ref{tab:exp-country} in Appendix). This further suggests that LLM-generated suggestions may be particularly helpful for participants less confident in English, supporting their ability to write constructively.
 
\section{Discussion}

\subsection{Promoting Constructive Discourse Online}
Through cross-cultural experiments with American and Indian participants, we demonstrate that LLMs can help people from different cultures write constructive comments on socially divisive issues like homophobia and Islamophobia. Our findings show that comments generated solely by an LLM or co-written with an LLM were perceived as more constructive than those written by humans alone. The LLM-generated comments also had significantly more constructive features than human-written comments. When participants used LLM assistance for commenting, it made their comments longer, more polite, positive, less toxic, and more readable, while enhancing argumentative features and preserving the original meaning. Overall, participants found the LLM helpful, were satisfied with the outputs as the LLM articulated their points clearly, and reduced the perceived difficulty of writing constructive comments. 

These findings align with prior research showing that people find LLM assistance valuable when countering hateful opinions~\cite{Mun2024} because LLMs are capable of rephrasing users' messages to convey respectful listening~\cite{argyle2023} and enhance receptiveness to opposing views~\cite{kambhatla2024}. Even recipients perceive LLM-generated arguments as stronger, more persuasive, and positive than those written by humans~\cite{Karinshak2023, argyle2023, kambhatla2024}. 

Our work contributes to this line of research in two key ways. First, we show that LLMs can help people from different cultures write constructive comments in response to contentious online threads on divisive issues. In contrast, most prior research has focused on LLMs’ capabilities in helping users craft argumentative essays on select topics or in private, one-on-one debates~\cite{Lee2022, Dang2023, Zhang2023, argyle2023}. Second, we draw from cross-cultural differences in argumentation to reveal potential misalignment between how humans and LLMs characterize constructive comments (Section~\ref{sec:misalign}). Additionally, we show that LLM's tendency to apply positive sentiment often misrepresent nuances in people's stances on divisive issues. 

While our findings suggest that LLMs have the potential to assist users in constructively engaging in online debates on divisive social issues, there are some caveats. To begin with, such prosocial interventions promoting constructive discourse are particularly beneficial for well-intentioned users, who may not realize toxicity in their writing or who, in heated moments, get ``emotionally triggered'' and unintentionally use offensive or slighted language~\cite{warner2024, kiskola2023user}. LLM-infused writing tools can encourage greater mindfulness during online conflicts, helping users avoid comments they might regret later~\cite{Sleeper-2013, Warner-2021}. 
Such tools can help shift some of the responsibility for addressing hateful interactions during conflicts onto the senders themselves, and could prove particularly useful in end-to-end encrypted platforms, such as WhatsApp groups, where hateful content flourishes due to limited moderation~\cite{saha2021short}. 

While such tools may not deter users who intentionally spread hateful rhetoric, 
they could empower bystander users to challenge problematic behavior by providing support to express disagreement constructively~\cite{Baughan2021, gurgun2023}. This could be especially valuable in situations, where users might hesitate to intervene due to the effort and emotional toll involved in participating in online debate without support.

\subsection{Human-AI Misalignment in Constructive Discourse}\label{sec:misalign}
Although our findings show that LLMs could help people write constructive comments on divisive issues, we found key differences between how humans and LLMs assess constructiveness. For instance, in our study, the LLM rated dialectical comments as constructive significantly more than humans did. The LLM considered politeness and balancing contrasting viewpoints, whereas humans prioritized logic and facts while assessing constructiveness. This misalignment is likely to affect how humans and LLMs approach constructive disagreement. For example, prior studies note that human-written counter-speech employs more specific strategies~\cite{mun-etal-2023}, such as providing counterexamples and stating facts, both of which are characteristics of logical argumentation. In contrast, counter-speech generated by GPT-4, Llama2, and Alpaca tend to be less specific, more polite, and broadly denounce hatefulness~\cite{mun-etal-2023, song2024assessing, phutane_cold_2025}, aligning more with the middle-ground approach of dialectical argumentation. Due to these differences, humans often perceive LLM-generated counter-speech as less convincing.

In our study, this misalignment often led participants to reject, regenerate, or edit the LLM's suggestions, resulting in comments that were significantly more negative and toxic. 
Research shows that LLMs tend to prioritize \textit{``surface-level''} lexical cues (e.g., joy, anger, fear, offensiveness) more than humans do while assessing sentiment~\cite{das2024under}. We observed similar effects in Human-AI written comments in our study. As a result, several participants who rejected or edited LLM-generated suggestions reported not liking the formal, moderate, and non-human sounding language. In line with these findings, \citet{Zhang2023} also found that users often perceive LLM-generated suggestions to be \textit{``robotic, monotonous, and repetitive''} while writing long argumentative essays.

We noticed that some participants felt frustrated when the LLM misrepresented their opinions, especially when their views were nuanced rather than outright polarizing (e.g., user supported LGBTQ communities but opposed legalizing same-sex marriage). While this suggests that LLMs may struggle to capture the subtleties and complexities of people's stances, it also raises important questions about how human-AI misalignment could impact deliberation on divisive issues. For example, prior studies show that different cultures and communities have different boundaries and tolerance for the language used online. Some users strongly value expressing opinions in an uncensored way and find it patronizing when comments are moderated to be positive~\cite{kiskola2023user}. Close friends often use curse words to reflect real-life rapport, yet algorithms might flag and remove these interactions as inappropriate~\cite{Shahid2023}. Similarly, in some communities (e.g., LGBTQ+ forums), toxic language is sometimes used to foster in-group solidarity through humor~\cite{warner2024}. It is possible that LLM suggestions could be less useful in such groups that are open to having more direct and uncensored dialogue on divisive issues. In such cases, it is crucial that the LLM preserves user autonomy and provides them with the choice to either rephrase or retain their writing. 


LLM's tendency to align with mainstream or dominant views and to generate responses that lack diversity in perspectives raise concerns about their ability to accurately reflect varied opinions~\cite{das2024under}. While we did not find evidence of homogenization in human-AI written comments, 
this might be because participants had to write their opinions first before prompting the LLM to rephrase their comments. The distinction between prompting LLMs to rephrase comments versus co-writing comments with auto-complete LLM suggestions holds significant design implications, as several studies show that auto-complete suggestions from LLMs shift people's viewpoints and homogenize their writing~\cite{Jakesch-2023, williams2024bias, agarwal2024}. Moreover, prior studies show that people fear reframing their messages with LLMs would undermine their agency and credibility, resulting in insincere, diluted, and morally compromising responses~\cite{Mun2024, Baughan2021}. For example, one of our participants reported accepting LLM-generated \textit{``neutral''} suggestions while responding to an Islamophobic thread because they suspected their views might be  \textit{``too biased to meet the appropriate criteria''} because they were Muslim. Thus, suggestions from LLM can lead to algorithmic conformity~\cite{liel2020if}, suppress minority viewpoints, and curb freedom of speech over time~\cite{feuston2020} if individuals and communities start adjusting their opinions to conform to what LLMs think as \textit{``constructive.''}  Therefore, a systematic investigation is needed to determine how suggestions in the form of auto-complete or rephrasing might affect homogenization in writing.


\subsection{Enabling Constructive Discourse Across Cultures}
Prior research in social psychology shows that people from individualistic cultures prefer logical arguments, whereas those from collectivist cultures favor dialectical arguments~\cite{norenzayan2002cultural, Peng1999, nisbett2001culture}. However, similar to Indian participants, American participants perceived dialectical comments as more constructive than the logical ones. Even both groups chose similar prompts to make their comments constructive and their comments had similar levels of linguistic features of constructiveness. This could be because existing work on cross-cultural differences in argumentation is based on either offline conflicts~\cite{ohbuchi1994cultural} or long-form argumentative essays~\cite{norenzayan2000rules, norenzayan2002cultural}, which may not directly apply to short-form online comments. For instance, Toulmin's logical argumentation framework includes six elements: claim, data, warrant, backing, qualifier, and rebuttal~\cite{hitchcock2006}. However, in our study, the LLM-generated, human-written, and HAI-written comments averaged 70-80 words in length, which may be too brief to capture all elements of argumentation. This brevity might have made it difficult for people to distinguish between logical and dialectical styles. 
In future research, we plan to conduct follow-up experiments using comments of varying lengths to examine if comment length influences preferences for logical versus dialectical comments on divisive issues.

Additionally, for culturally grounded writing tasks, LLM-generated suggestions have been shown to lead Indian participants to adopt Western writing styles~\cite{agarwal2024}. In our study, American and Indian participants responded to different threads relevant to their cultural contexts, which prevented a direct comparison of their comments to determine if cultural homogenization occurred when co-writing constructive comments with the LLM. Therefore, in future research, we plan to investigate comments from Indian and American participants written in response to the same threads that are relevant to both cultural contexts. 
To address cultural homogenization from biased LLMs, techniques such as anthropological prompting~\cite{alkhamissi2024}, cultural prompting~\cite{tao2024}, self-pluralism~\cite{xu2024self}, modular pluralism~\cite{feng2024modular}, or value-pluralistic design~\cite{sorensen2024value}---which have been shown to increase cultural alignment in LLM-generated responses---are necessary to design HAI-collaborative systems to support prosocial discourse online on a global scale.

\subsection{Ethical Consideration and Feasibility of Facilitating Constructive Discourse}

Apart from the benefits of promoting constructive discourse, we need to critically think about the potential biases and abuses stemming from such systems. For instance, LLMs have been shown to produce responses with covert cultural harms in seemingly neutral language that are unlikely to be detected by existing methods~\cite{dammu2024they}. Since we only used existing sentiment and toxicity analyses tools, we may not have uncovered such covert biases in HAI-written comments. However, a manual review of the LLM-generated comments from Phase 1 revealed statements like \textit{``...its [Islam's] damaging treatment towards women, LGBTQ+ individuals, and non-Muslims...''}---which was generated in response to an Islamophobic thread, but from the stance of opposing Islamophobia. If LLMs perpetuate such covert biases in the form of \textit{``constructive''} discourse while taking the stance of supporting marginalized communities, it would strengthen existing stereotypes and potentially harm marginalized individuals. 


\citet{carstens2024ai} have also critiqued such AI tools due to their simplified view of online civility. They argue that apart from input-output, these systems need to be evaluated within existing social inequalities and hierarchies. Because, the argumentative norms facilitated by these tools would privilege expressions from highly educated people, who usually have better training in writing such arguments, while disregarding linguistic and cultural variations.

Additionally, existing research on the role of LLMs in promoting prosocial discourse has primarily relied on crowd evaluations~\cite{kambhatla2024} or assessed receptiveness only among small group of discussants~\cite{argyle2023, Tessler-2024}. Although research shows that promoting civil dialogue can actually enhance user engagement on the platform~\cite{liu2024measuring}, there is limited understanding of how such interventions would work in large online communities, where the discussions can be viewed by anyone. \citet{cho2023} have noticed that that even when LLM-based intervention encourages deep listening, empathy, and critical thinking, it struggles to instill respectful and cooperative attitude among people. Therefore, future work should look into: how many users need to write their comments constructively to affect the course of a divisive online thread? How would this affect users' engagement and subsequent conversation quality?
\subsection{Limitations and Future Work}
Our work has some limitations. First, we focused on two divisive social issues—Islamophobia and homophobia—and participants from two geographies, India and the US, as proxies for different cultures. Therefore, the findings may not be generalizable to other regions or social issues. Moreover, we used only GPT-4 for writing constructive comments. Although GPT-4's polite and broad, middle-ground approach aligns with that of other models like Llama2 and Alapaca~\cite{mun-etal-2023, song2024assessing}, future work should investigate the quality of constructive comments generated by different language models.

Second, when verifying the argumentation styles in LLM-generated comments, we noticed many comments were either wrongly labeled or failed to achieve  consensus among annotators. This led us to regenerate several comments. Given that humans and LLMs emphasize different features when assessing constructiveness---despite both preferring dialectical argumentation---these inconsistencies raise question about the argumentation quality of LLM-generated comments and their alignment with humans' notion of constructiveness. Future work should involve expert evaluations to rigorously assess the argumentation quality of LLM-generated comments.

Third, while prior research indicates that annotators' identities influence how they perceive online content~\cite{sap-etal-2022}, our analysis did not account for how participants' identities (e.g., gender and religion) might have shaped their perceptions and evaluations of constructiveness in the comments. To mitigate this, we ensured participants reviewed comment pairs, where both comments were from the same stance. However, future research should deeply look into how identity influences perceptions of constructive comments in online debates on divisive topics.

Fourth, we recruited Indian and American crowd workers on Prolific as representatives of collectivist and individualist cultures, respectively. Although we did not administer any instrument to verify cultural differences among Indian and American Prolific workers, prior research shows significant cultural differences between these two groups~\cite{agarwal2024}. Since we did not find any cultural difference in our study despite both countries having lots of cultural diversity, future research should explicitly control for the participants' cultural context. 

Finally, although our results provide initial evidence that LLMs can help people write more constructive comments, we assessed audiences' perceptions of constructiveness rather than the actual impact of these comments---such as whether they shift opinions or foster common ground on divisive issues.
We also did not investigate how many constructive comments are needed within an online debate to meaningfully influence the tone of the conversation. Although this is not a limitation of this work, it represents a critical next step that we aim to address in future research.

\section{Conclusion}
Through controlled experiments involving 600 participants from India and the US, our study takes an important first step toward understanding how LLMs can help people from different cultures write constructive comments on divisive issues. Our study reveals potential misalignment between how humans and LLMs assess constructiveness, prioritizing different aspects of argumentation styles. Despite these differences, LLMs are able to help people write their opinions more constructively than they could do on their own. We noticed that although LLMs could capture people's points well without homogenizing their opinions, it often misrepresented people's views on divisive issues. These findings highlight the promise and limitations of using LLMs in cross-cultural, value-sensitive contexts, and underscore the need for socio-technical systems that better align with diverse human perspectives on socially divisive issues.

\begin{acks}
    We thank Infosys and Cornell Global AI initiative for supporting this work. We also thank anonymous reviewers for their constructive and respectful feedback on our work.
\end{acks}

\bibliographystyle{ACM-Reference-Format}
\bibliography{00_References}

\appendix

\section{Appendix}


\begin{table}[ht]
\caption{Title of the original Reddit threads that participants reviewed in our study.}
\label{tab:thread_titles}
\resizebox{0.8\columnwidth}{!}{
\begin{tabular}{|l|l|l|}
\hline
\multicolumn{1}{|c|}{\textbf{Topic}} & \multicolumn{1}{c|}{\textbf{India}}                                                                                                                      & \multicolumn{1}{c|}{\textbf{US}}                                                                                                                                      \\ \hline
Homophobia                           & \begin{tabular}[c]{@{}l@{}}1. Should gay marriage be \\ legalised in India?\\ 2. Which gay man, without a \\ uterus, has a menstrual cycle?\end{tabular} & \begin{tabular}[c]{@{}l@{}}1. Lib thinks republicans are out \\ to get them because they’re gay lol\\ 2. Speaker Mike Johnson’s \\ Obsession With Gay Sex\end{tabular} \\ \hline
Islamophobia                         & \begin{tabular}[c]{@{}l@{}}1. Why my otherwise liberal \\ family has a problem with Islam\\ 2. Islamic Takeover of India \\ by 2047\end{tabular}         & \begin{tabular}[c]{@{}l@{}}1. Islamophobia is a great thing and \\ i’m tired of being called racist for it\\ 2. I am Islamophobic\end{tabular}                         \\ \hline
\end{tabular}}
\end{table}

\begin{figure}[ht]
    \centering
    \setlength{\fboxsep}{0pt}%
    \setlength{\fboxrule}{1pt}%
    \fbox{\includegraphics[width=\linewidth]{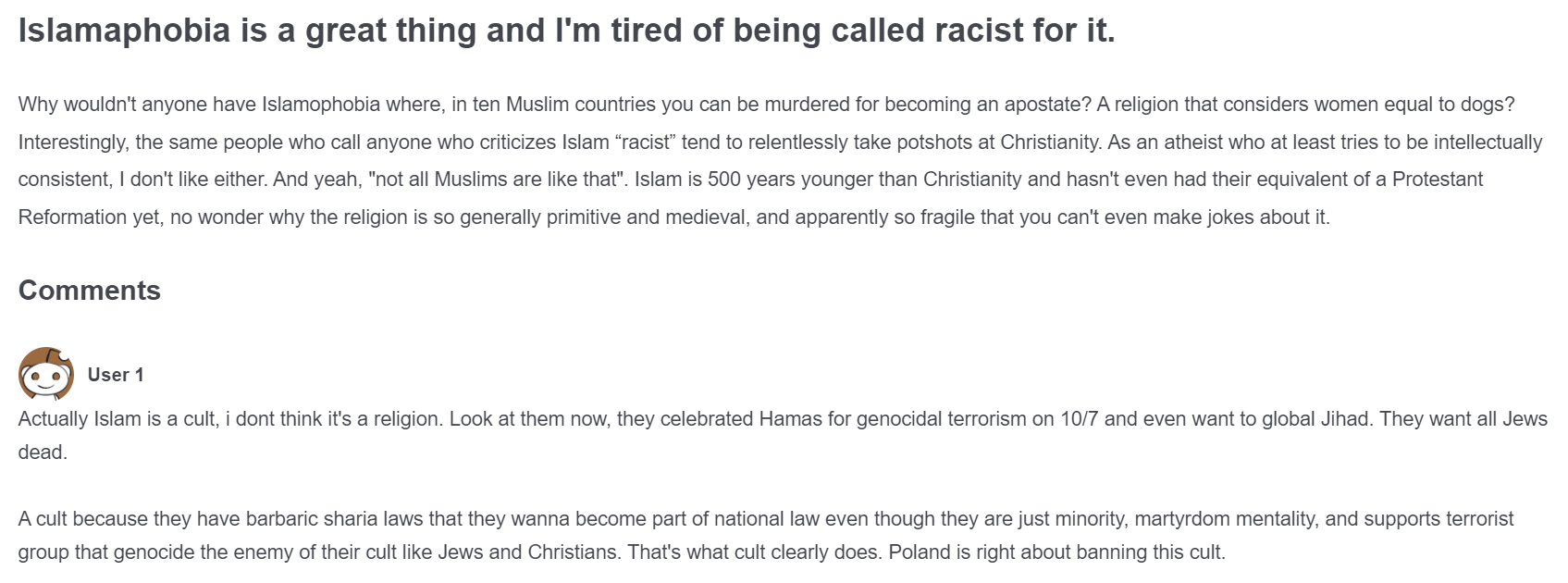}}
    \caption{A segment of the Islamophobic thread shown in the context of the US.}
    \Description{Original post: Islamophobia is a great thing and i'm tired of being called racist for it. 
    
    Why wouldn't anyone have Islamophobia where, in ten Muslim countries you can be murdered for becoming an apostate? A religion that considers women equal to dogs? Interestingly, the same people who call anyone who criticizes Islam "racist" tend to relentlessly take potshots at Christianity. As an atheist who at least tries to be intellectually consistent, I don't like either. And yeah, "not all Muslims are like that". Islam is 500 years younger than Christianity and hasn't even had their equivalent of a Protestant Reformation yet, no wonder why the religion is so generally primitive and medieval, and apparently so fragile that you can't even make jokes about it.

    Comments

    User 1: Actually Islam is a cult, i dont think it's a religion. Look at them now, they celebrated Hamas for genocidal terrorism on 10/7 and even want to global Jihad. They want all Jews dead.
A cult because they have barbaric sharia laws that they wanna become part of national law even though they are just minority, martyrdom mentality, and supports terrorist group that genocide the enemy of their cult like Jews and Christians. That's what cult clearly does. Poland is right about banning this cult.}
    \label{fig:us-islamophobic-post}
\end{figure}


\parabold{Model Hyper-parameters}\label{hyp-parameter} To generate constructive comments on incendiary topics using GPT-4, we used the following hyper-parameters from prior studies on opinionated and argumentative writing with AI~\cite{minaLee-2022, Jakesch-2023, Karinshak2023}. High values of these parameters increase randomness and produce non-repetitive outputs. We experimented with three different values of frequency penalty to generate three comments for each of the 32 cases. These cases were derived from the combination of 2 demographics $\times$ 2 issues $\times$ 2 threads $\times$ 2 stances $\times$ 2 argumentation styles.
\begin{itemize}
    \item Sampling temperature = 0.8
    \item Top P (nucleus sampling) = 1
    \item Presence penalty = 0
    \item Frequency penalty = \{0, 0.5, 1\}
\end{itemize}

\begin{table}[ht]
\caption{Prompts for GPT-4 to generate constructive comments with different argumentation styles.}
\label{tab:comment_gen}
\resizebox{\columnwidth}{!}{
\begin{tabular}{|l|l|}
\hline
\textbf{Constructive comments with logical argumentation}                                                                                                                                                                                                                                                                                                                                                                                                                                                                                                           & \textbf{Constructive comments with dialectical argumentation}                                                                                                                                                                                                                                                                                                                                                                                                                                                                                                    \\ \hline
\begin{tabular}[c]{@{}l@{}}Consider the following Reddit thread:\\ \textless{}insert Reddit thread\textgreater\\ \\ You are assisting an \textless{}American/ Indian\textgreater user, who think \\ \textless{}insert stance\textgreater of \textless{}homophobia/ Islamophobia\textgreater{}. Help the\\ user write a constructive comment in response to this\\ thread.\\ \\ The comment must use analytic rules and formal logic to\\ write evidence based arguments. The comment must be\\ assertive, direct and stay within 100 words.\end{tabular} & \begin{tabular}[c]{@{}l@{}}Consider the following Reddit thread:\\ \textless{}insert Reddit thread\textgreater\\ \\ You are assisting an \textless{}American/ Indian\textgreater user, who think \\ \textless{}insert stance\textgreater of \textless{}homophobia/ Islamophobia\textgreater{}. Help the \\ user write a constructive comment in response to this\\ thread.\\ \\ The comment must be indirect, and apply dialectical \\ argument. The comment must take a middle\\ ground and stay within 100 words.\end{tabular} \\ \hline
\end{tabular}
}
\end{table}

\begin{table}[ht]
\caption{Comparison of linguistic features of constructiveness between LLM-generated logical and dialectical comments using parametric and non-parametric tests, with Bonferroni corrections applied.}
\label{tab:constructive-feature-ai-gen}
\resizebox{0.9\columnwidth}{!}{
\begingroup

\begin{tabular}{|c|cc|c|}
\hline
\multirow{2}{*}{\textbf{Features}} & \multicolumn{2}{c|}{\textbf{Average values}}                                 & \multirow{2}{*}{\textbf{Statistics}}           \\ \cline{2-3}
                                   & \multicolumn{1}{c|}{\textbf{Logical comment}} & \textbf{Dialectical comment} &                                                \\ \hline
Length (words)                     & \multicolumn{1}{c|}{84}                       & 81                           & -                                              \\ \hline
Readability                        & \multicolumn{1}{c|}{13.5}                     & 14.6                         & t(65)=-2.88, p\textless{}0.01, Cohen's d=0.7   \\ \hline
Politeness                         & \multicolumn{1}{c|}{10.15}                    & 12.76                        & -                                              \\ \hline
Named entity                       & \multicolumn{1}{c|}{2.88}                     & 1.74                         & -                                              \\ \hline
Discourse connectives              & \multicolumn{1}{c|}{10.24}                    & 10.03                        & -                                              \\ \hline
Stance adverbials                  & \multicolumn{1}{c|}{2.10}                     & 0.44                         & U=989.5, Z=-5.28, p\textless{}0.000001, r=0.64 \\ \hline
Reasoning verbs \& modals          & \multicolumn{1}{c|}{6.74}                     & 7.29                         & -                                              \\ \hline
Root clauses                       & \multicolumn{1}{c|}{6.42}                     & 5.15                         & U=865.5, Z=-3.65, p\textless{}0.001, r=0.44    \\ \hline
\end{tabular}
\endgroup
}
\end{table}


\begin{table}[ht]
\caption{Perceived characteristics of constructive comments reported by Indian and American Participants}
\label{tab:char-in-US}
\resizebox{0.6\columnwidth}{!}{
\begin{tabular}{|l|c|c|}
\hline
\textbf{Characteristics}                    & \multicolumn{1}{l|}{\textbf{India (\%)}} & \multicolumn{1}{l|}{\textbf{US (\%)}} \\ \hline
More relevant to conversation               & 20                                  & 17                                \\ \hline
Balances different viewpoints better        & 18                                  & 21                                \\ \hline
More polite and respectful language         & 18                                  & 19                                \\ \hline
Takes a better solution-oriented approach   & 23                                  & 23                                \\ \hline
Better logic and facts to support arguments & 21                                  & 20                                \\ \hline
\end{tabular}}
\end{table}

\begin{table}[ht]
\caption{Multiple pairwise Mann-Whitney tests with Bonferroni corrections comparing features of constructiveness across HAI vs. Human, Human vs. AI, and HAI vs. AI comments.}
\label{tab:constructive-feature}
\resizebox{0.7\columnwidth}{!}{
\begin{tabular}{|c|c|c|c|}
\hline
\textbf{\begin{tabular}[c]{@{}c@{}}Constructive\\ characteristics\end{tabular}} & \textbf{HAI vs. AI}                                                                     & \textbf{Human vs. AI}                                                                   & \textbf{HAI vs. Human}                                                                  \\ \hline
Length                                                                          & \begin{tabular}[c]{@{}c@{}}U=5106, Z=5.52,\\ p\textless{}0.000001, r=0.43\end{tabular} & \begin{tabular}[c]{@{}c@{}}U=4944, Z=6.62,\\ p\textless{}0.000001, r=0.52\end{tabular} & -                                                                                      \\ \hline
Readability score                                                               & -                                                                                      & \begin{tabular}[c]{@{}c@{}}U=5546, Z=8.55,\\ p\textless{}0.000001, r=0.68\end{tabular} & \begin{tabular}[c]{@{}c@{}}U=7843, Z=8.64,\\ p\textless{}0.000001, r=0.63\end{tabular} \\ \hline
Politeness markers                                                              & -                                                                                      & -                                                                                      & -                                                                                      \\ \hline
Named entities                                                                  & -                                                                                      & -                                                                                      & -                                                                                      \\ \hline
Discourse connectives                                                           & \begin{tabular}[c]{@{}c@{}}\textcolor{changes}{U=4356, Z=3.11,}\\ \textcolor{changes}{p\textless{}0.001, r=0.24}\end{tabular}    & \begin{tabular}[c]{@{}c@{}}\textcolor{changes}{U=4151, Z=3.70,}\\ \textcolor{changes}{p\textless{}0.001, r=0.29}\end{tabular}    & -                                                                                      \\ \hline
Stance adverbials                                                               & -                                                                                      & \begin{tabular}[c]{@{}c@{}}\textcolor{changes}{U=4070, Z=3.70,}\\ \textcolor{changes}{p\textless{}0.001, r=0.29}\end{tabular}    & -                                                                                      \\ \hline
Reasoning verbs and modals                                                      & -                                                                                      & \begin{tabular}[c]{@{}c@{}}\textcolor{changes}{U=4173, Z=3.79,}\\ \textcolor{changes}{p\textless{}0.001, r=0.30}\end{tabular}    & -                                                                                      \\ \hline
Root clauses                                                                    & \begin{tabular}[c]{@{}c@{}}\textcolor{changes}{U=5595, Z=7.33,}\\ \textcolor{changes}{p\textless{}0.000001, r=0.57}\end{tabular} & \begin{tabular}[c]{@{}c@{}}\textcolor{changes}{U=4866, Z=6.26,}\\ \textcolor{changes}{p\textless{}0.000001, r=0.50}\end{tabular} & -                                                                                      \\ \hline
\end{tabular}}
\end{table}

\begin{table}[ht]
\caption{Results of independent ANCOVA tests with Bonferroni corrections to examine the effect of comment type on argumentative features while controlling for comment length.}
\label{tab:ancova-test}
\resizebox{\columnwidth}{!}{
\begingroup

\begin{tabular}{|l|l|l|}
\hline
\textbf{Features}          & \multicolumn{1}{c|}{\textbf{Effect of comment type}} & \multicolumn{1}{c|}{\textbf{Effect of comment length}} \\ \hline
Discourse connectives      & F(2, 255)=9.57, p\textless{}0.0001, $\eta^{2}$=0.04           & F(1, 255)=160.13, p\textless{}0.000001, $\eta^{2}$=0.37         \\ \hline
Stance adverbials          & F(2, 255)=14.24, p\textless{}0.00001, $\eta^{2}$=0.1          & \multicolumn{1}{c|}{-}                                 \\ \hline
Reasoning verbs and modals & F(2, 255)=7.91, p\textless{}0.001, $\eta^{2}$=0.05            & F(1, 255)=64.08, p\textless{}0.000001, $\eta^{2}$=0.19          \\ \hline
Root clauses               & F(2, 255)=42.01, p\textless{}0.000001, $\eta^{2}$=0.21        & F(1, 255)=70.86, p\textless{}0.000001, $\eta^{2}$=0.17          \\ \hline
\end{tabular}
\endgroup}
\end{table}

\begin{table}[ht]
\caption{Results from multiple Wilcoxon signed rank tests with Bonferroni corrections to compare the characteristics between initial human-written and final HAI co-written comments in the test group.}
\label{tab:cultural-arg-result}
\resizebox{0.7\columnwidth}{!}{
\begin{tabular}{|l|l|}
\hline
\textbf{Characteristics}   & \textbf{Statistics}                            \\ \hline
Length                     & W=260, Z=-6.94, p\textless{}0.000001, r=0.53   \\ \hline
Discourse connectives      & \textcolor{changes}{W=393, Z=-5.70, p\textless{}0.000001, r=0.43}    \\ \hline
Stance adverbials          & \multicolumn{1}{c|}{\textcolor{changes}{-}} \\ \hline
Reasoning verbs and modals & \textcolor{changes}{W=229, Z=-6.49, p\textless{}0.000001, r=0.49}   \\ \hline
Root clauses               & \textcolor{changes}{W=109.5, Z=-7.29, p\textless{}0.000001, r=0.55}   \\ \hline
Readability score          & W=6.5, Z=-8.06, p\textless{}0.000001, r=0.61   \\ \hline
Named entity               & \multicolumn{1}{c|}{-}                         \\ \hline
Politeness                 & W=914, Z=-3.91, p\textless{}0.001, r=0.30      \\ \hline
Sentiment                  & W=415, Z=-6.24, p\textless{}0.000001, r=0.47   \\ \hline
Toxicity                   & W=3627, Z=7.25, p\textless{}0.000001, r=0.55   \\ \hline
\end{tabular}}
\end{table}

\begin{table}[ht]
\caption{Results of independent ANCOVA tests with Bonferroni corrections to examine the effect of LLM's suggestions on argumentative features while controlling for comment length.}
\label{tab:ancova-test-hai}
\resizebox{\columnwidth}{!}{
\begingroup

\begin{tabular}{|l|l|l|}
\hline
\textbf{Features}          & \multicolumn{1}{c|}{\textbf{Effect of comment type}} & \multicolumn{1}{c|}{\textbf{Effect of comment length}} \\ \hline
Discourse connectives      & F(1, 171)=55.64, p\textless{}0.000001, $\eta^{2}$=0.16        & F(1, 171)=124.83, p\textless{}0.000001, $\eta^{2}$=0.36         \\ \hline
Stance adverbials          & \multicolumn{1}{c|}{-}                               & F(1, 171)=11.19, p\textless{}0.001, $\eta^{2}$=0.06             \\ \hline
Reasoning verbs and modals & F(1, 171)=70.08, p\textless{}0.000001, $\eta^{2}$=0.21        & F(1, 171)=97.12, p\textless{}0.000001, $\eta^{2}$=0.29          \\ \hline
Root clauses               & F(1, 171)=167.7, p\textless{}0.000001, $\eta^{2}$=0.35        & F(1, 171)=140.6, p\textless{}0.000001, $\eta^{2}$=0.29          \\ \hline
\end{tabular}
\endgroup}
\end{table}

\begin{table}[ht]
\caption{Linguistic features of constructiveness in comments written by Indian and American Participants}
\label{tab:char-control-test}
\resizebox{\columnwidth}{!}{
\begin{tabular}{c|c|c|c|c|c|c|c|c|c|}
\cline{2-10}
\multicolumn{1}{l|}{}                          & \multicolumn{1}{l|}{Country} & \multicolumn{1}{l|}{\begin{tabular}[c]{@{}l@{}}Length\\ (words)\end{tabular}} & \multicolumn{1}{l|}{\begin{tabular}[c]{@{}l@{}}Discourse\\ connective\end{tabular}} & \multicolumn{1}{l|}{\begin{tabular}[c]{@{}l@{}}Stance\\ adverbial\end{tabular}} & \multicolumn{1}{l|}{\begin{tabular}[c]{@{}l@{}}Reasoning\\ verb \& modal\end{tabular}} & \multicolumn{1}{l|}{\begin{tabular}[c]{@{}l@{}}Root\\ clause\end{tabular}} & \multicolumn{1}{l|}{Politeness} & \multicolumn{1}{l|}{\begin{tabular}[c]{@{}l@{}}Named\\ entity\end{tabular}} & \multicolumn{1}{l|}{Readability} \\ \hline
\multicolumn{1}{|c|}{\multirow{2}{*}{Control}} & India                        & 75                                                                            & 3.75                                                                                & 0.88                                                                            & 2.50                                                                                   & 2.65                                                                       & 15.50                           & 2.69                                                                        & 8.83                             \\ \cline{2-10} 
\multicolumn{1}{|c|}{}                         & US                          & 68                                                                            & 3.35                                                                                & 0.82                                                                            & 2.49                                                                                   & 2.59                                                                       & 15.64                           & 1.41                                                                        & 8.95                             \\ \hline
\multicolumn{1}{|c|}{\multirow{2}{*}{Test}}    & India                        & 74                                                                            & 4.35                                                                                & 1.47                                                                            & 2.65                                                                                   & 2.57                                                                       & 13.10                           & 1.76                                                                        & 14.42                            \\ \cline{2-10} 
\multicolumn{1}{|c|}{}                         & US                          & 70                                                                            & 3.96                                                                                & 1.51                                                                            & 2.61                                                                                   & 2.55                                                                       & 12.04                           & 1.63                                                                        & 13.41                            \\ \hline
\end{tabular}}
\end{table}

\begin{table}[ht]
\caption{Mann-Whitney tests with Bonferroni corrections comparing Indian and American participants' experiences of writing constructive comments using LLM.}
\label{tab:exp-country}
\begin{tabular}{|l|cc|c|}
\hline
\multicolumn{1}{|c|}{\multirow{2}{*}{\textbf{Experience}}} & \multicolumn{2}{c|}{\textbf{Average value}}       & \multirow{2}{*}{\textbf{Statistics}}                             \\ \cline{2-3}
\multicolumn{1}{|c|}{}                                     & \multicolumn{1}{c|}{\textbf{India}} & \textbf{US} &                                                                  \\ \hline
Satisfaction                                               & \multicolumn{1}{c|}{4.41}           & 3.68        & \multicolumn{1}{l|}{U=700.5, Z=4.09, p\textless{}0.0001, r=0.41} \\ \hline
Ownership                                                  & \multicolumn{1}{c|}{4.33}           & 3.49        & \multicolumn{1}{l|}{U=696.5, Z=4.08, p\textless{}0.0001, r=0.41} \\ \hline
Difficulty                                                 & \multicolumn{1}{c|}{1.98}           & 2.45        & -                                                                \\ \hline
\end{tabular}
\end{table}

\end{document}